\documentclass[a4paper,10pt]{article}

\usepackage{jheppub} 

\usepackage[T1]{fontenc} 

\usepackage{amsmath}            
\usepackage{amssymb}
\usepackage{graphicx}  
\usepackage{booktabs}
\usepackage{comment}
\usepackage{hyperref}
\usepackage{subfigure}

\newcommand{\cF}{{\cal F}}
\newcommand{\NLL}{\text{NLL}}
\newcommand{\NNLL}{\text{NNLL}}
\newcommand{\FNLL}{\mathcal{F}_{\rm NLL}}
\newcommand{\FNNLL}{\mathcal{F}_{\rm NNLL}}

\newcommand{\Vsc}[1]{V_{\rm sc}(\{\tilde p \},#1)}

\newcommand{\dF}[1]{\delta\mathcal{F}_{\rm #1}}
\newcommand{\ptilde}{\{\tilde p \}}

\title{\boldmath The Sudakov radiator for jet observables and the
  soft physical coupling}

\author[a]{Andrea Banfi,}
\author[a]{Basem Kamal El-Menoufi,}
\author[b]{and Pier Francesco Monni}

\affiliation[a]{Department of Physics and Astronomy, University of
  Sussex, Sussex House, Brighton, BN1 9RH, UK}
\affiliation[b]{CERN, Theoretical Physics Department, CH-1211 Geneva 23,
Switzerland}

\emailAdd{a.banfi@sussex.ac.uk}
\emailAdd{b.elmenoufi@sussex.ac.uk}
\emailAdd{pier.monni@cern.ch}

\abstract{ We present a procedure to calculate the Sudakov radiator
  for a generic recursive infrared and collinear (rIRC) safe
  observable in two-scale problems. We give closed formulae for the
  radiator at next-to-next-to-leading-logarithmic (NNLL) accuracy,
  which completes the general NNLL resummation for this class of
  observables in the {\tt ARES} method for processes with two emitters
  at the Born level. As a byproduct, we define a physical coupling in
  the soft limit, and we provide an explicit expression for its
  relation to the $\overline{\rm MS}$ coupling up to
  ${\cal O}(\alpha_s^3)$. This physical coupling constitutes one of
  the ingredients for a NNLL accurate parton shower algorithm.  As an
  application we obtain analytic NNLL results, of which several are
  new, for all angularities $\tau_x$ defined with respect to both the
  thrust axis and the winner-take-all axis, and for the moments of
  energy-energy correlation $FC_x$ in $e^+e^-$ annihilation. For the
  latter observables we find that, for some values of $x$, an accurate
  prediction of the peak of the differential distribution requires a
  simultaneous resummation of the logarithmic terms originating from
  the two-jet limit and at the Sudakov shoulder.}

\preprint{CERN-TH-2018-179}

\begin{document}

\maketitle

\flushbottom

\section{Introduction}
\label{sec:intro}


Distributions in event shapes and jet resolution parameters,
collectively jet observables, are among the most studied QCD
observables. Since they are continuous measures of the hadronic
energy-momentum flow in jet events at colliders, they constitute a
powerful probe of the dynamics of strong interactions, from high
scales where fixed-order perturbative calculations can be applied,
down to low scales where the yet unexplained phenomenon of
hadronisation plays a decisive role.

Jet observables play a major role in measurements of the QCD coupling
$\alpha_s$, and in testing non-perturbative hadronisation models (see
e.g.\ Ref.~\cite{Patrignani:2016xqp} and references therein).
The study of jet observables also led to important advances in the
understanding of all-order properties of QCD radiation, which lead to
the discovery of the so-called non-global
logarithms~\cite{NG1,NG2,Banfi:2002hw}.
Distributions in jet observables can be computed at fixed order in QCD
perturbation theory. Such calculations have reached
next-to-next-leading order (NNLO) accuracy for a number of relevant
QCD processes. In particular, for $e^+e^-$ annihilation, NNLO
corrections to three-jet production have been computed in
refs.~\cite{GehrmannDeRidder:2007hr,GehrmannDeRidder:2008ug,Weinzierl:2008iv,Weinzierl:2009ms,DelDuca:2016ily}.

While fixed-order calculations provide a reliable tool to describe jet
observables in the region where their values are large, the bulk of
data lies in a region where multiple soft-collinear emissions give
rise to large logarithms of the jet observable at all orders in
perturbation theory.
To be precise, given a generic jet observable, let us consider its
cumulative distribution $\Sigma(v)$, the fraction of events such that
the observable's value is less than $v$. This quantity exhibits
logarithmic terms as large as $\alpha_s^n L^{2n}$, where $L=-\ln v$ and
$n$ is the order in QCD perturbation theory. Resumming those large
logarithms means reorganising $\ln\Sigma$ in such a way that it can be
written as
$\alpha_s g_1(\alpha_s L)+g_2(\alpha_s L)+\alpha_s g_3(\alpha_s
L)+\dots$,
where $g_1(\alpha_s L)$ resums the so-called leading logarithmic (LL)
contributions, $\alpha_s^n L^{n+1}$, $g_2(\alpha_s L)$ the NLL ones,
$\alpha_s^n L^n$, $g_3(\alpha_s L)$ the NNLL ones,
$\alpha_s^n L^{n-1}$, and so on.

Next-to-leading logarithmic (NLL) resummations, that include all terms
${\cal O}(\alpha_s^nL^n)$ in the logarithm of cumulative distributions,
have been available for many years for specific
observables~\cite{CSS,CTTW,Bonciani:2003nt,thr_res,mh_ee,cpar_res,DLMSBroadPT}.
Nowadays, NLL resummation for jet observables that have the properties
of recursive infrared and collinear (rIRC) safety and continuous
globalness~\cite{Banfi:2004yd,Banfi:2003je} is a solved problem. The
general solution is based on a semi-numerical approach developed for
$e^+e^-$ event-shapes and jet rates in Ref.~\cite{Banfi:2001bz}, and
later extended to any suitable jet observable in any QCD hard
process~\cite{Banfi:2004yd,Banfi:2003je}.  The method is implemented
in the computer program {\tt CAESAR}~\cite{Banfi:2004yd}, that also
verifies whether a given observable is rIRC safe and continuously
global.
This led to a first systematic study of event shapes in hadronic dijet
production at NLL accuracy matched to next-to-leading order (NLO)
results at hadron colliders~\cite{Banfi:2004nk,Banfi:2010xy}.

NLL predictions have a sizeable theoretical uncertainty. Given the
precision of current experiments, theoretical accuracy for
resummations should aim at NNLL, and in some cases beyond.
Most NNLL resummations are observable specific, and rely on the
properties of the observable to achieve resummation through
factorisation theorems that separate different kinematical
configurations (e.g. hard, soft, collinear), and appropriate
renormalisation group equations based on the fact that physical
distributions do not depend on the unphysical scales that need to be
introduced to achieve such separation.
Such approaches made it possible to obtain full
next-to-next-to-leading logarithmic (NNLL) predictions for a number of
global $e^+e^-$ event shapes such as thrust
$1-T$~\cite{Becher:2008cf,Abbate:2010xh,Monni:2011gb}, heavy jet mass
$\rho_H$~\cite{Chien:2010kc}, jet broadenings
$B_T$,~$B_W$~\cite{Becher:2012qc}, $C$-parameter~\cite{Hoang:2014wka},
energy-energy correlation
(EEC)~\cite{deFlorian:2004mp,Tulipant:2017ybb,Moult:2018jzp}, heavy
hemisphere groomed mass~\cite{Frye:2016okc}, and
angularities~\cite{Procura:2018zpn}. Among the above examples, for
$1-T$, $\rho_H$, $C$-parameter, and EEC, all N$^3$LL corrections are
also known, except for the four-loop cusp anomalous dimension, that
has been computed numerically more recently~\cite{Moch:2018wjh}.
Jet observables have been resummed at NNLL accuracy also in deep
inelastic scattering~\cite{Kang:2013nha,Kang:2013wca,Kang:2013lga}.
For hadronic collisions, full NNLL resummations are available for
processes where a colour singlet is produced at Born level,
specifically for a boson's transverse
momentum~\cite{Bozzi:2005wk,Becher:2010tm} and
$\phi^*$~\cite{Banfi:2011dx}, the beam
thrust~\cite{Stewart:2010pd,Berger:2010xi}, transverse
thrust~\cite{Becher:2015lmy}, and the leading jet's transverse
momentum~\cite{Becher:2012qa,Becher:2013xia,Banfi:2012jm,Stewart:2013faa},
and for heavy quark pair's transverse
momentum~\cite{Zhu:2012ts,Catani:2014qha}. For an arbitrary number of
legs, a NNLL accurate resummation is available for the $N$-jettiness
variable~\cite{Jouttenus:2011wh}.
Very recently, resummations for the boson's transverse momenta and
$\phi^*$ have been pushed to N$^3$LL
accuracy~\cite{Bizon:2017rah,Bizon:2018foh,Chen:2018pzu}.

Despite these remarkable results, most jet observables are beyond the
scope of factorisation theorems. This is especially true for those
observables which cannot be expressed in terms of simple analytic
functions of momenta, e.g.\ event shapes like the thrust major, or the
two-jet rate in the Durham algorithm. For rIRC safe observables, it is
possible to achieve NNLL accuracy by means of the semi-numerical
method {\tt ARES} (Automated Resummer for Event Shapes), developed for
$e^+e^-$ event shapes in Ref.~\cite{Banfi:2014sua}, and later extended
to jet rates~\cite{Banfi:2016zlc}. These publications focused on NNLL
corrections induced by resolved real radiation. There, the
cancellation of infrared singularities between unresolved real and
virtual corrections to the Born process was parametrised in an
observable dependent Sudakov form factor called ``radiator'', that was
extracted from existing calculations. This made it possible to study
only observables that scale like powers of the transverse momentum or
the invariant mass of the jet, which constitute a vast set of the
phenomenologically relevant observables. This led to the first
resummations for complicated observables such as the thrust major and
the two-jet rate with various jet algorithms in
$e^+e^-$~\cite{Banfi:2014sua,Banfi:2016zlc}.

In this paper we complete the last analytic ingredient necessary to
have a fully general formula for the resummation of rIRC safe jet
observables at NNLL in processes with two hard legs at the Born level.
This makes it possible to handle all known observables of this type in
a single framework, and hence paves the way to systematic
phenomenological applications. We formulate the Sudakov radiator at
all orders for a generic rIRC jet observable in QCD, and we explicitly
compute it at NNLL accuracy. Our calculation of the Sudakov radiator
must be then supplemented with the finite contributions coming from
real radiation that are computed as in
refs.~\cite{Banfi:2014sua,Banfi:2016zlc}.

The paper is organised as follows. In section~\ref{sec:ares} we give
the formulation of a general procedure for the resummation of jet
observables with the {\tt ARES} formalism. There we define the main
object of our paper, the Sudakov radiator, which we compute at NNLL
accuracy in section~\ref{sec:rad-NNLL}. The computation of the
radiator leads to a definition of the physical coupling for soft
radiation at higher orders. This generalises the scheme of
Ref.~\cite{Catani:1990rr}, and constitutes an ingredient for future
NNLL accurate parton shower algorithms. In that same section, we
present a new formulation of the NNLL correction $\dF{correl}$
introduced in Ref.~\cite{Banfi:2014sua}, that we redefined in order to
make sure that the Sudakov radiator can be computed analytically for
an arbitrary rIRC safe jet observable. In section~\ref{sec:FC}, we
apply our method to angularities and moments of energy-energy
correlation in $e^+e^-$ annihilation, and outline briefly the main
features of their phenomenology at present and future
colliders. Section~\ref{sec:the-end} contains our conclusions.

\section{Resummation in the {\tt ARES} formalism}
\label{sec:ares}

In this section we summarise the structure of the NNLL resummed cross
section for a generic rIRC observable in $e^+e^-$ annihilation. We
first set up the relevant notation and kinematics, and then we move on
to derive the resummed cross section up to NNLL accuracy.

\subsection{Kinematics and notation}
\label{sec:kinematics}
Let $V(\{\tilde p\},k_1,\dots,k_n)$ be a generic continuously global,
rIRC safe final-state observable, a function of all final-state
momenta.\footnote{The actual inputs of final-state observables are
  hadron momenta. However, it is well known that using parton momenta
  gives distributions that differ from the measured ones by
  corrections suppressed by powers of the typical hard scale of the
  process, in this case the centre-of-mass energy of the $e^+e^-$
  collision.} Here $\{\tilde p\}$ denotes $\{\tilde p_1,\tilde p_2\}$,
which are the quark-antiquark pair initiating the process after all
radiation has been emitted, and $k_1,\dots,k_n$ are the momenta of the
additional radiation. At Born level $V(\{\tilde p\})=0$.
In order to parametrise the radiation momenta $k_i$, we introduce the
following Sudakov decomposition
\begin{align}
\label{eq:sudakov}
k_i^\mu = z_i^{(1)} p^{\mu}_1 + z_i^{(2)} p^{\mu}_2 + \kappa_i^\mu,
\end{align}
where $p^{\mu}_1, p^{\mu}_2$ are two light-like reference vectors, and
$\kappa_i^\mu$ is a space-like four-vector and its magnitude is
denoted by $k_{ti} \equiv \sqrt{-\kappa_i^2}$. The on-shell condition
$k_i^2=0$ implies that $z_i^{(1)} z_i^{(2)} 2 (p_1p_2) = k_{ti}^2$.
The choice of the reference momenta $p_1$ and $p_2$ is arbitrary, and
determines the mapping between the Sudakov variables
$\{z_i^{(1)},z_i^{(2)},\kappa_i\}$ and the actual final-state momenta
$\{\tilde p_1,\tilde p_2,k_1,\dots,k_n\}$.
A particular choice is therefore motivated by computational
convenience. In our case, we observe that the product of the squared
amplitude and phase space for an emission $k$ collinear to either
$\tilde p_1$ or $\tilde p_2$ is proportional to
\begin{equation}
  \label{eq:ktilde-squared}
[dk] M^2(k) \propto \frac{d\tilde k_t^ 2}{\tilde k_t^2},\qquad {\rm where}\qquad{\tilde k}_t^2 = \frac{2(\tilde p_1 k) 2(\tilde p_2 k)}{2(\tilde p_1
   \tilde p_2)}.
\end{equation}
We choose the reference vectors  $p_1$ and $p_2$ such that 
\begin{equation}
\label{eq:ktmeasure}
\frac{d k_t^ 2}{ k_t^2}=\frac{d\tilde k_t^ 2}{\tilde k_t^2}
\end{equation}
up to corrections that vanish as a power of $k_t$ in the limit
$k_t\to 0$.
One possible solution, adopted in Ref.~\cite{Banfi:2004yd}, is to
choose the two reference vectors $p_1$ and $p_2$ in
Eq.~\eqref{eq:sudakov} to be the momenta of the emitter of parton
$k_i$ and the corresponding spectator. The precise definition of
emitter and spectator requires specifying an ordering of insertion of
the emissions in the event.
A natural way of reconstructing the kinematics is to insert the
emissions as follows:
\begin{itemize}
\item Start with the two-parton event consisting of the initial
  $q\bar q$ pair, without any additional emissions. These momenta
  define the initial $p_1$ and $p_2$ vectors.
\item Consider a set of emissions parametrised by the triplet
  $\{z_i^{(1)},z_i^{(2)},\kappa_i\}$. At this stage the kinematics is
  not uniquely determined, as we did not specify the reference vectors
  of Eq.~\eqref{eq:sudakov}. 
\item Associate emissions for which $z_i^{(1)} > z_i^{(2)}$ to leg
  $\tilde p_1$, and emissions for which $z_i^{(2)} > z_i^{(1)}$ to leg
  $\tilde p_2$.  For each emission $k_i$, define its
  rapidity $\eta_i$ with respect to the emitting leg as
\begin{align}
\label{eq:rapidity}
  \eta_i^{(1)} &= \frac{1}{2}\ln \frac{z_i^{(1)}}{z_i^{(2)}}\,,\quad {\rm if}\quad z_i^{(1)} > z_i^{(2)}
 \,,\\
\eta_i^{(2)} &= \frac{1}{2}\ln \frac{z_i^{(2)}}{z_i^{(1)}}\,,\quad {\rm if}\quad z_i^{(1)} < z_i^{(2)}\,.
\end{align}
\item For each leg $\ell$, insert the emissions into the initial event
  starting from the one at smaller $\eta^{(\ell)}$. The reference
  momentum $p_\ell$ in Eq.~\eqref{eq:sudakov} represents the emitter
  and takes the transverse recoil. The longitudinal recoil is shared
  between the emitter and a spectator momentum (i.e.\ the remaining
  reference vector in Eq.~\eqref{eq:sudakov}). However, for the
  purpose of the analytical method presented in this paper, one can
  safely neglect the longitudinal recoil of the spectator (which is
  proportional to $k_{ti}^2$) that would otherwise give rise to
  regular terms in the cross section. For instance, for an emission
  $k$ emitted off leg $1$, and parametrised by Eq.~\eqref{eq:sudakov},
  we have
\begin{equation}
p_1^\mu \to (1-z^{(1)})p_1^\mu - \kappa^\mu\,,\qquad
p_2^\mu \to p_2^\mu.
\end{equation}
The resulting momenta $p_1$ and $p_2$, after the emission, will be
massless up to ${\cal O}(k_{ti}^2)$ corrections.
\item Update the reference momenta, and proceed with the insertion of
  emissions at progressively larger $\eta_i^{(\ell)}$.
\end{itemize}
This procedure guarantees the validity of Eq.~\eqref{eq:ktmeasure},
and implies that the reference vectors $p_1$ and $p_2$ are different
for each emission~\cite{Banfi:2014sua}. 

Before proceeding, we stress that the ordering chosen for the
insertion of the emissions in the event has nothing to with the way
the triplets $\{z_i^{(1)},z_i^{(2)},\kappa_i\}$ are produced. For
instance, when computing the resummation via a Monte Carlo algorithm,
it is convenient to generate this triplet according to an ordering in
the observable's value (see,
e.g.~\cite{Banfi:2004yd,Banfi:2014sua}). This ordering is however
unrelated to the ordering with which the emissions are inserted, which
follows the angular ordering arguments outlined above. This is
essential to specify a correct recoil scheme that preserves the simple
factorised form of QCD matrix elements.

\subsection{General structure of NNLL resummation}
\label{sec:resummation}
Any rIRC safe observable $V(\{\tilde p\},k_1,\dots,k_n)$, can be
parametrised in the following way for a single soft and collinear
emission $k$ collinear to each leg $\ell$
\begin{align}\label{eq:sc-behaviour}
    V(\{\tilde p\},k) \simeq V_{\text{sc}}(k) \equiv \sum_{\ell=1}^2 d_\ell \left(\frac{k_t^{(\ell)}}{Q}\right)^a
  e^{- b_{\ell} \eta^{(\ell)}} g_\ell(\phi^{(\ell)})\Theta(\eta^{(\ell)}) \, .
\end{align}
Here $k_t^{(\ell)},\eta^{(\ell)},\phi^{(\ell)}$ are the transverse
momentum, rapidity and azimuth of $k$ with respect to the emitter
$p_\ell$ as defined in the previous section, whereas $a,b_\ell,d_\ell$
are constants.\footnote{IRC safety implies $a>0$ and $b_\ell>
  -a$.
  While $b_\ell,d_\ell$ and $g_\ell$ can be different for each leg,
  continuous globalness implies $a$ has to be the same for all legs.}
The scale $Q$ represents a typical hard scale of the process, in our
case the centre-of-mass energy of the $e^+e^-$ collision.

Our aim is to resum large logarithms in the cumulative distribution
$\Sigma(v)$, the fraction of events for which
$V(\{\tilde p\},k_1,\dots,k_n) < v$, in the region $v\ll 1$. This is
given by
\begin{align}\label{eq:cumbasic}
\Sigma(v) \equiv \frac{1}{\sigma}\int_0^v dv' \frac{d \sigma(v')}{d
  v'} = \mathcal{H}(Q) \sum_{n=0}^\infty \frac{1}{n!} \int \prod_{i=1}^n [dk_i] \mathcal{M}^2(k_1,...,k_n) \Theta( v - V(\{\tilde p\},k_1,\dots,k_n)  )\,,
\end{align} 
where $\mathcal{H}(Q)$ includes all virtual corrections to the Born
process (normalised to the total cross section $\sigma$) and
$\mathcal{M}^2(k_1,...,k_n)$ is the amplitude squared for $n$ real
emissions. The Lorentz-invariant phase-space in $d=4-2\epsilon$
dimensions is denoted by $[dk]$ and defined as
\begin{align}
\label{eq:dk}
[dk] \equiv \frac{d^d k}{(2\pi)^{d-1}} \delta(k^2)\Theta(k^0).
\end{align} 
Given the Sudakov decomposition of any four-momentum $k$ of (squared) invariant
mass $k^2=m^2$ as in Eq.~(\ref{eq:sudakov}), the measure $d^d k$ can
be expressed as
\begin{align}\label{eq:PSmassive}
 d^d k = (p_1 p_2) dz^{(1)} dz^{(2)} d^{2-2\epsilon}k_t = \frac{d y}{2} dm^2 d^{2-2\epsilon}k_t, \qquad y \equiv \frac12 \ln\left(\frac{z^{(1)}}{z^{(2)}}\right)\,.
\end{align}
The variable $y$ is the rapidity of $k$ (with respect to some reference light-like directions $p_1$ and $p_2$), and for real emissions, i.e.\ $k^2=0$, it is bounded by\footnote{Strictly, one should use $\sqrt{2(p_1 p_2)}$ instead of the centre-of-mass energy of the $e^+e^-$ collision $Q$. However, for the emissions of relevance in this paper, the two scales coincide, up to corrections that vanish as a power of $k_t$.}
\begin{equation}
  \label{eq:y-bound}
  |y|<\ln\left(\frac{Q}{k_t}\right)\,.
\end{equation}
It is immediate to link $y$ to the rapidity of a massless emission $k$
with respect to a given leg $\ell$. In fact, $\eta^{(1)}=y$ for $y>0$,
and $\eta^{(2)}=-y$ for $y<0$. This implies that the phase space
$[dk]$ in Eq.~\eqref{eq:dk} can be written as follows
\begin{equation}
  \label{eq:dk-final}
  [dk] = \sum_{\ell=1}^2 \frac{d\eta^{(\ell)}}{2}\Theta(\eta^{(\ell)})\frac{d^{2-2\epsilon}k_t}{(2\pi)^{3-2\epsilon}} \,.
\end{equation}
So far, all momenta have to be considered in $d$-dimensions, because
dimensional regularisation is needed to regulate the IRC divergences
present both in $\mathcal{H}(Q)$ and in the real radiation.

The virtual corrections $\mathcal{H}(Q)$ can be expressed
as~\cite{Laenen:2000ij,Dixon:2008gr}
\begin{align}
\label{eq:virtuals}
\mathcal{H}(Q)  = C(\alpha_s(Q)) & \exp\left\{-\int\frac{d^d k}{(2\pi)^d}\, w(m^2,k_t^2+m^2;\epsilon)\,\Theta\left(\frac{1}{2}\ln\left(\frac{Q^2}{k_t^2+m^2}\right) -|y|\right)\Theta(Q-k_t)\right\}\times \nonumber \\ 
& \times \exp\left\{ - \sum_{\ell=1}^2\int^{Q^2} \frac{d k^2}{k^2} \gamma_\ell(\alpha_s(k,\epsilon)) \right\}\ \ .
\end{align}
The virtual corrections are parametrised in three objects. First,
$w(m^2,k_t^2+m^2;\epsilon)$ denotes the soft web~\cite{FT,NAET} of
total momentum $k$ and squared invariant mass $m^2$, in
$d=4-2\epsilon$ dimensions. The web is obtained by considering the
Feynman graphs with two eikonal lines that cannot be further
decomposed into subgraphs by cutting each eikonal line once. For
example, at lowest order in perturbation theory we find
\begin{equation}
  \label{eq:web-1}
  w^{(1)}(m^2,m^2+k_t^2;\epsilon) = (4\pi)^2 \frac{2C_\ell}{k_t^2} \mu_R^{2\epsilon}\alpha_s(\mu_R) \delta(m^2)\,,
\end{equation}
where $C_\ell=C_F$ in the case of quarks, and $C_\ell=C_A$ in the case
of emitting gluons.  Note that the web does not depend on the rapidity
$y$, due to the properties of eikonal Feynman rules. Remarkably, at
each order in perturbation theory, the web has a finite limit for
$\epsilon\to 0$, which we will simply denote by
$w(m^2,m^2+k_t^2)$. Second, the function
$\gamma_\ell(\alpha_s(k,\epsilon))$ coincides, up to an overall sign
change, with the coefficient of the $\delta(1-x)$ term of the
regularised splitting functions $P_{qq}(x)$ and $P_{gg}(x)$, according
to whether leg $\ell$ is a quark or a gluon, respectively. The strong
coupling $\alpha_s(k,\epsilon)$ is defined as the solution of the
$d$-dimensional renormalisation group equation:
\begin{equation}
\mu_R^2\frac{d\alpha_s}{d\mu_R^2} = -\epsilon\, \alpha_s + \beta^{\rm (d=4)}(\alpha_s),
\end{equation}
where $\beta^{\rm (d=4)}$ is the beta function in four dimensions,
given by the following expansion
\begin{equation}
  \label{eq:beta-function}
  \beta^{\rm (d=4)}(\alpha_s) = - \alpha_s^2 \sum_{n=0}^\infty \beta_n \alpha_s^n\,.
\end{equation}
In our representation of $\mathcal{H}(Q)$, the upper integration bound
for the $k_t$-integral of the web is set by the centre-of-mass energy
$Q$, and the upper bound for the rapidity integral to
$|y|<\ln(Q/\sqrt{k_t^2+m^2})$. Finally, the overall quantity
$C(\alpha_s(Q))$ is a multiplicative constant that is obtained by
matching Eq.~\eqref{eq:virtuals} at each order in perturbation theory
to the quark or gluon form factor computed in the $\overline{\rm MS}$
scheme.

In order to proceed, we need to define a procedure to cancel the IRC
singularities in Eq.~\eqref{eq:virtuals} against those in the real
emissions. This can be done by introducing a resolution parameter that
is engineered in such a way to divide the real radiation into a {\it
  resolved} set and an {\it unresolved} one. The idea behind this
procedure is to handle the unresolved part of the radiation
analytically, and hence cancel the divergences against the virtual
corrections. The cancellation is performed in a manner that the resolved
contributions could subsequently be computed numerically in $d=4$ dimensions.
The resolution parameter is defined through its action on the soft
and/or hard-collinear contributions to the squared amplitudes, as
outlined below.

We start by considering soft radiation. The soft squared amplitudes
for $n$ emissions, denoted hereafter as $M_{\rm s}^2(k_1,\dots, k_n)$,
can be iteratively reorganised as follows
\begin{align}
\nonumber
M_{\rm s}^2(k_1) &\equiv \tilde{M}_{\rm s}^2(k_1) \\\nonumber 
M_{\rm s}^2(k_1, k_2) &= \tilde{M}_{\rm s}^2(k_1) \tilde{M}_{\rm s}^2(k_2) + \tilde{M}_{\rm s}^2(k_1,k_2) \\ 
M_{\rm s}^2(k_1, k_2, k_3) &= \tilde{M}_{\rm s}^2(k_1) \tilde{M}_{\rm s}^2(k_2)
                     \tilde{M}_{\rm s}^2(k_3) + \left(\tilde{M}_{\rm s}^2(k_1)
                     \tilde{M}_{\rm s}^2(k_2,k_3) + \text{perm.}\right) +
                     \tilde{M}_{\rm s}^2(k_1,k_2,k_3)\, \nonumber\\
\label{eq:clusters}
\vdots 
\end{align}
The quantities $\tilde{M}_{\rm s}^2(k_1, \dots,k_n)$ represent the
correlated portion of the $n$-emission soft amplitude squared,
together with its virtual corrections.\footnote{Note that the  $\tilde{M}_{\rm s}^2(k_1, \dots,k_n)$ are not in general positive definite, in that they are defined as differences of squared matrix elements.} This is strongly suppressed
unless all emissions $k_1,\dots,k_n$ are close in angle. We refer to
the latter as {\em soft correlated blocks} and they play a dominant
role in constructing the {\em webs}, which are the building objects of
the Sudakov radiator to be defined below.  Each correlated block
admits a perturbative expansion in $\alpha_s$ due to virtual
corrections, hence
\begin{equation}
\tilde{M}_{\rm s}^2(k_1, \dots,k_n) = \tilde{M}_{\rm s, 0}^{2}(k_1,
\dots,k_n) + \frac{\alpha_s(\mu_R)}{2\pi}\tilde{M}_{\rm s, 1}^{2}(k_1,
\dots,k_n) + \dots
\end{equation}
For instance, at tree level, the squared matrix element for the
emission of a single soft gluon is given by
\begin{equation}
\tilde{M}_{\rm s}^2(k) \simeq \tilde{M}_{\rm s, 0}^2(k)  =  16 \pi\, C_\ell\, \mu_R^{2\epsilon} \frac{\alpha_s(\mu_R)}{k_t^2},
\end{equation}
while at one-loop order one has~\cite{Catani:2000pi}
\begin{equation}
\tilde{M}_{\rm s, 1}^2(k) = - \tilde{M}_{\rm s, 0}^2(k)C_A
    \frac{1}{\epsilon^2}\frac{\Gamma^4(1-\epsilon)\Gamma^3(1+\epsilon)}{\Gamma^2(1-2
  \epsilon)\Gamma(1+2\epsilon)} \left(\frac{4\pi\mu_R^2}{k_t^2}\right)^\epsilon.
\end{equation}
In the following the coupling will be always renormalised in the
$\overline{\rm MS}$ scheme, i.e. we replace
\begin{equation}
\mu_R^{2\epsilon}\alpha_s(\mu_R)\to
\mu_R^{2\epsilon}\alpha_s(\mu_R)\frac{e^{\epsilon
    \gamma_E}}{(4\pi)^\epsilon}
\left(1 - \frac{\beta_0}{\epsilon}\alpha_s(\mu_R)+\dots\right)\,,
\end{equation}
where $\beta_0$ is the first coefficient of the beta function in four dimensions of Eq.~(\ref{eq:beta-function}), given by
\begin{equation}
  \label{eq:beta0}
  \beta_0 = \frac{11 C_A - 2 n_f}{12 \pi}\,. 
\end{equation}
The tree-level correlated block with two emissions
$\tilde{M}_{\rm s, 0}^2(k_1,k_2)$ is reported in
Appendix~\ref{sec:2ps}, and will be useful later.
This decomposition is particularly convenient to define a logarithmic
counting. Each correlated block $\tilde{M}_{\rm s}^2(k_1, \dots,k_n)$
will contribute to $\ln\Sigma(v)$ at most with a factor
$\alpha_s^n\ln^{n+1} (v)$, with $n$ powers of $\ln(v)$ coming from the
soft singularities and an extra power from the only collinear
singularity.

The definition of the resolution parameter proceeds as follows. First,
we define a {\em clustering algorithm} that combines together the
momenta of all particles emitted according to each correlated block
$\tilde{M}_{\rm s}^2(k_1, \dots,k_n)$.
For example consider the simple case of two emissions, the clustering
is assigned as follows
\begin{equation}
 M _{\rm s}^2(k_1, k_2) =   \underbrace{\tilde{M}_{\rm s}^2(k_1) \tilde{M}_{\rm s}^2(k_2) }_{\text{ two clusters } (k_1,k_2)}\qquad + \underbrace{\tilde{M}_{\rm s}^2(k_1,k_2)}_{\text{a single cluster }k_{\text{clust.}} =k_1+k_2 } \ \ .
\end{equation}
The property of rIRC safety~\cite{Banfi:2004yd} implies that all
particles in a cluster are both close in angle and have commensurate
transverse momenta. This allows one to evaluate the QCD running
couplings of each cluster at the transverse momenta of the
corresponding emissions. This procedure allows us to absorb all
logarithms of $\mu_R/k_{ti}$ into the running of the coupling. As a
consequence of this procedure, for rIRC safe observables, using the
decomposition~\eqref{eq:clusters}, every correlated block
$\tilde{M}_{\rm s}^2(k_1, \dots,k_n)$ (when combined with the
corresponding virtual corrections) will contribute to $\ln\Sigma(v)$
with terms of order $\alpha_s^m\ln^{m+2-n}(v)$ for $m\geq n$. This
allows us to build a logarithmic counting at the level of the squared
amplitude, which defines which contributions must be considered at a
given logarithmic order.

In order to proceed with the calculation of $\Sigma(v)$, we then
choose a resolution parameter $\delta \ll 1$ such that all clusters of
total momentum $k_{\text{clust.}}$ satisfying
\begin{align}\label{eq:resobsbasic}
V_{\text{sc}}(k_{\text{clust.}}) < \delta v \,, \ 
\end{align}
are labelled as {\it unresolved}.
This choice guarantees that one is able to compute analytically the
contribution of unresolved emissions for an arbitrary rIRC safe
observable.
For this class of observables, the unresolved clusters can be
neglected from the $\Theta$ function in Eq.~\eqref{eq:cumbasic} since
they do not contribute to the observable $V$ up to corrections
suppressed by powers of $\delta^p v$, with $p$ being a positive
parameter.

The above definition of the resolution parameter allows us to
exponentiate the contribution of unresolved soft blocks. 
From the decomposition of Eq.~\eqref{eq:clusters}, it is
straightforward to connect the correlated blocks
$\tilde{M}_{\rm s}^2(k_1, \dots,k_n)$ to the webs introduced in
Eq.~\eqref{eq:virtuals}. In fact
\begin{equation}
  \label{eq:web-connect}
  w(m^2,k_t^2+m^2;\epsilon)=\sum_{n=1}^\infty S(n)\int \left(\prod_{i=1}^n
  [dk_i]\right) \tilde{M}_{\rm s}^2(k_1,...,k_n) (2\pi)^d\delta^{(d)}(k-\sum_i
  k_i)\,,
\end{equation}
where the factor $S(n)$ represents the multiplicity coefficient for
each soft final state (quarks or gluons). For instance, for $n$
identical gluons, $S(n)=1/n!$.  Therefore, the contribution of an
arbitrary number of soft clusters (and no hard-collinear clusters)
gives rise to the following exponential factor
\begin{align}
\label{eq:unresolved_soft}
\exp \left\{ \int^Q\frac{d^d k}{(2\pi)^{d}}  w(m^2,k_t^2+m^2;\epsilon) \Theta(\delta v - V_{\rm sc}(k))\right\}\,.
\end{align}
Eq.~\eqref{eq:unresolved_soft} can be promptly combined with the
virtual corrections in~\eqref{eq:virtuals} to give
\begin{align}
&\mathcal{H}(Q) \exp \left\{ \int^Q\frac{d^d k}{(2\pi)^{d}}
  w(m^2,k_t^2+m^2;\epsilon) \Theta(\delta v - V_{\rm
  sc}(k))\right\}\notag\\
\label{eq:soft_cancellation}
& = C(\alpha_s(Q)) e^{-R_{\rm s}(\delta v)}\exp\left\{- \sum_{\ell=1}^2\int^{Q^2} \frac{d k^2}{k^2} \gamma_\ell(\alpha_s(k,\epsilon))\right\}\,,
\end{align}
where we defined the soft radiator $R_{\rm s}$ as
\begin{align} 
\label{eq:soft_radiator}
R_{\rm s}(v) = \int^Q\frac{d^4
  k}{(2\pi)^{4}} w(m^2,k_t^2+m^2) \Theta(V_{\rm sc}(k)-v)\,,
\end{align}
and we took the four-dimensional limit of the web (and the relative
integration measure) since the integral is now finite.

The next step is to handle the remaining hard-collinear divergences
present in the integral over $\gamma$ in
Eq.~\eqref{eq:soft_cancellation}. Unlike the case of soft radiation,
the exponentiation of the unresolved hard-collinear emissions is more
delicate in that every hard-collinear emission could potentially
change the colour charge felt by subsequent radiation. However, the
treatment of hard-collinear radiation is much simplified by observing
that, owing to rIRC safety, only a fixed number of hard-collinear
emissions is to be considered at a given logarithmic order. Therefore,
instead of proceeding as in the soft case, it is convenient to start
from the integral over the anomalous dimension $\gamma$ in the virtual
corrections~\eqref{eq:soft_cancellation} and split it into two pieces
at the collinear scale of the resolution variable, that is found by
setting the rapidity $\eta^{(\ell)}$ to its maximum
(i.e. $\ln(Q/k_t))$ in Eq.~\eqref{eq:sc-behaviour}, which yields
$V_{\rm sc}(k) \sim k_t^{a+b_\ell}$. Inspired by this, the integral
over $\gamma$ then can be split at $k=v^{1/(a+b_\ell)}Q$ and becomes
\begin{align}
\int^{Q^2} \frac{d k^2}{k^2} \gamma_\ell(\alpha_s(k,\epsilon)) =\int^{Q^2}_{Q^2 v^{
  \frac{2}{a+b_\ell}}}\frac{d k^2}{k^2} \gamma_\ell(\alpha_s(k)) +\int^{Q^2 v^{ \frac{2}{a+b_\ell}}}_0\frac{d k^2}{k^2} \gamma_\ell(\alpha_s(k,\epsilon))\,.
\end{align}
Next, we expand the exponential of the second integral in the
r.h.s. of the above equation considering only a fixed number of terms
in its expansion as
\begin{equation}
\label{eq:hc_sudakov}
\exp\left\{ -\int^{Q^2 v^{ \frac{2}{a+b_\ell}}}_0\frac{d k^2}{k^2}
  \gamma_\ell(\alpha_s(k,\epsilon)) \right\} = 1- \int^{Q^2 v^{
    \frac{2}{a+b_\ell}}}_0\frac{d k^2}{k^2}
\frac{\alpha_s(k,\epsilon)}{2\pi}\gamma^{(0)}_\ell + {\cal O}(\alpha_s^2(Q v^{
    \frac{1}{a+b_\ell}})),
\end{equation}
where, in our case of emitting quarks, the leading-order anomalous
dimension is given by
\begin{equation}
\label{eq:gamma0}
\gamma^{(0)}_\ell = -\frac{3}{2}\,C_F.
\end{equation}
The first non-trivial order in the expansion must be included at NNLL,
the second at N$^3$LL (together with the squared of the first), and so
forth.
The divergences of these terms will cancel order-by-order in
perturbation theory against those of the hard-collinear emissions in
the real radiation.

The last step to obtain a NNLL expression for $\Sigma(v)$ is to handle
the squared matrix element for real emissions ${\cal M}^2$ in
Eq.~\eqref{eq:cumbasic}. At NLL accuracy, rIRC safety ensures that
resolved radiation contains no hard-collinear emissions, and the real
matrix element squared is approximated by its soft approximation
$M_{\rm s}^2$.  Moreover, the squared amplitude at this order reduces
to the product of $n$ independent, soft-collinear emission
probabilities. In fact, $[dk]M_{\rm s,0}^2(k) \simeq [dk]M^2_{\rm sc}(k)$, where~\cite{Banfi:2014sua}
\begin{equation}
  \label{eq:matrix-element}
[dk]M^2_{\rm sc}(k)\equiv\sum_{\ell=1,2} 2 C_\ell \frac{\alpha_s(k^{(\ell)}_{t})}{\pi}\frac{dk^{(\ell)}_{t}}{k^{(\ell)}_{t}} 
 d\eta^{(\ell)}\, \Theta\left(\ln\left(\frac{Q}{k^{(\ell)}_{t}}\right)-\eta^{(\ell)}\right) \Theta(\eta^{(\ell)})
\frac{d\phi^{(\ell)}}{2\pi}\,,
\end{equation}
and $C_\ell$ the colour factor of leg $\ell$, $C_F$ for a quark and $C_A$
for a gluon.
In order to achieve NNLL accuracy, it is sufficient to correct the
products of independently emitted single-particle clusters with the
insertion of a single tree-level correlated cluster of two soft and
collinear emissions $\tilde{M}_{\rm s, 0}^2(k_a,k_b)$, and of the
one-loop correction to the single-emission cluster
$\tilde{M}_{\rm s, 1}^2(k)$.

Moreover, beyond NLL, a finite number of hard-collinear emissions must
be considered. In particular, at NNLL, it is sufficient to allow one
single emission to be hard and collinear.  When combined with
Eq.~\eqref{eq:hc_sudakov}, this leads to a finite, logarithmically
enhanced, left over, as it will be shown shortly. In such
configurations, at NNLL, the remaining soft radiation consists of an
arbitrary number of single-emission clusters.

With the above decomposition, the NNLL resummed cross section 
$\Sigma(v)$ of Eq.~\eqref{eq:cumbasic} takes the form
\begin{align}\label{eq:cumbasic_NNLL}
&\Sigma_{\rm NNLL}(v) = C(\alpha_s(Q)) e^{-R_{\rm s}(\delta v)}\exp\left\{- \sum_{\ell=1}^2\int^{Q^2}_{Q^2 v^{
  \frac{2}{a+b_\ell}}}\frac{d k^2}{k^2}
  \gamma_\ell(\alpha_s(k))\right\}\notag\\
& \times \bigg\{ \sum_{n=0}^\infty \frac{1}{n!} \int\prod_{i=1}^n
  [dk_i] M_{\rm s}^2(k_1,...,k_n) \Theta( v - V(\{\tilde
  p\},k_1,\dots,k_n)) \prod_{\rm clust.}\Theta\left(V_{\rm
  sc}(k_{\rm clust.} )- \delta v\right) \notag\\
& + \sum_{n=0}^\infty \frac{1}{n!}\int\prod_{i=1}^n
  [dk_i] M_{\rm sc}^2(k_i) \Theta\left(V_{\rm
  sc}(k_{i} )- \delta v\right)\times \sum_{\ell=1}^2\bigg[\int [dk_{\rm hc}] M_{\rm
  hc,\ell}^2(k_{\rm hc})\Theta( v - V(\{\tilde
  p\},k_1,\dots,k_n,k_{\rm hc})) \notag\\
&-  \int^{Q^2 v^{
    \frac{2}{a+b_\ell}}}_0\frac{d k^2}{k^2}
\frac{\alpha_s(k,\epsilon)}{2\pi}\gamma^{(0)}_\ell \Theta( v - V(\{\tilde
  p\},k_1,\dots,k_n)) \bigg] \bigg\},
\end{align} 
where the squared amplitude $M_{\rm s}^2(k_1,...,k_n)$ is approximated
by
\begin{equation}
  \begin{split}
M_{\rm s}^2(k_1,...,k_n)  &  \prod_{\rm clust.}\Theta\left(V_{\rm
  sc}(k_{\rm clust.} )- \delta v\right)  \simeq \prod_{i=1}^n M_{\rm sc}^2(k_i) \Theta\left(V_{\rm
  sc}(k_i )- \delta v\right) \\ &
+ \sum_{a>b}\prod_{\substack{i=1\\ i\neq a,b}}^{n}M_{\rm sc}^2(k_i) \Theta\left(V_{\rm
  sc}(k_i )- \delta v\right) \tilde{M}_{\rm s, 0}^2(k_a,k_b)\Theta\left(V_{\rm
  sc}(k_a +k_b)- \delta v\right) \\ & + \frac{\alpha_s(\mu_R)}{2\pi}\sum_{a}\prod_{\substack{i=1\\ i\neq a}}^{n}M_{\rm sc}^2(k_i) \Theta\left(V_{\rm
  sc}(k_i )- \delta v\right) \tilde{M}_{\rm s, 1}^2(k_a) \Theta\left(V_{\rm
  sc}(k_a )- \delta v\right)\,,
  \end{split}
\end{equation}
where $V_{\rm sc}(k_a +k_b)$ is defined as in
Eq.~(\ref{eq:sc-behaviour}), and
$k_t^{(\ell)}, \eta^{(\ell)}, \phi^{(\ell)}$ are the transverse
momentum, rapidity and azimuth of the four-vector $k_a+k_b$ with
respect to leg $\ell$.  The cancellation of infrared and collinear
divergences in the first term of Eq.~\eqref{eq:cumbasic_NNLL} can be
easily handled with a simple subtraction scheme as outlined in
Ref.~\cite{Banfi:2014sua}, which allows for a numerical evaluation in
$d=4$ dimensions. The cancellation of collinear singularities in the
second term still requires the use of dimensional regularisation. In
order to make the second term suitable for a numerical evaluation, we
add and subtract the following counter-term
\begin{align}
C(\alpha_s(Q))& e^{-R_{\rm s}(\delta v)}\exp\left\{- \sum_{\ell=1}^2\int^{Q^2}_{Q^2 v^{
  \frac{2}{a+b_\ell}}}\frac{d k^2}{k^2}
  \gamma_\ell(\alpha_s(k,\epsilon))\right\}\notag\\
&\times\sum_{n=0}^\infty \frac{1}{n!}\int\prod_{i=1}^n
  [dk_i] M_{\rm sc}^2(k_i) \Theta\left(V_{\rm
  sc}(k_{i} )- \delta v\right) \notag\\
&\times \sum_{\ell=1}^2\int [dk_{\rm hc}] M_{\rm
  hc,\ell}^2(k_{\rm hc}) \Theta( v - V(\{\tilde
  p\},k_1,\dots,k_n))\Theta(v-V_{\rm sc}(k_{\rm hc})),
\end{align}
and recast Eq.~\eqref{eq:cumbasic_NNLL} as follows 
\begin{align}\label{eq:cumbasic_NNLL_start}
&\Sigma(v) = C(\alpha_s(Q)) e^{-R_{\rm s}(\delta v)}\exp\left\{- \sum_{\ell=1}^2\int^{Q^2}_{Q^2 v^{
  \frac{2}{a+b_\ell}}}\frac{d k^2}{k^2}
  \gamma_\ell(\alpha_s(k,\epsilon))\right\}\notag\\
& \times \bigg\{ \sum_{n=0}^\infty \frac{1}{n!} \int\prod_{i=1}^n
  [dk_i] M_{\rm s}^2(k_1,...,k_n) \Theta( v - V(\{\tilde
  p\},k_1,\dots,k_n)) \prod_{\rm clust.}\Theta\left(V_{\rm
  sc}(k_{\rm clust.} )- \delta v\right) \notag\\
& + \sum_{n=0}^\infty \frac{1}{n!}\int\prod_{i=1}^n
  [dk_i] M_{\rm sc}^2(k_i) \Theta\left(V_{\rm
  sc}(k_{i} )- \delta v\right) \notag\\
&\times \sum_{\ell=1}^2\bigg[\int [dk_{\rm hc}] M_{\rm
  hc,\ell}^2(k_{\rm hc})\bigg(\Theta( v - V(\{\tilde
  p\},k_1,\dots,k_n,k_{\rm hc})) - \Theta( v - V(\{\tilde
  p\},k_1,\dots,k_n))\Theta(v-V_{\rm sc}(k_{\rm hc}))\bigg)\notag\\
&+  \bigg(\int [dk_{\rm hc}] M_{\rm
  hc,\ell}^2(k_{\rm hc})\Theta(v-V_{\rm sc}(k_{\rm hc})) -\int^{Q^2 v^{
    \frac{2}{a+b_\ell}}}_0\frac{d k^2}{k^2}
\frac{\alpha_s(k,\epsilon)}{2\pi}\gamma^{(0)}_\ell \bigg)\Theta( v - V(\{\tilde
  p\},k_1,\dots,k_n)) \bigg]\bigg\}.
\end{align} 
The integral in round brackets of the last line of the above equation
can be evaluated analytically as follows. For each leg $\ell=1,2$, we
expand the $\Theta(v-V_{\rm sc}(k_{\rm hc})) $ function in the last
line of Eq.~\eqref{eq:cumbasic_NNLL_start} as 
\begin{align}
\label{eq:theta_expansions}
\Theta(v-V_{\rm sc}(k_{\rm hc})) &= \Theta\left(v-d_\ell \left(\frac{k_t^{(\ell)}}{Q}\right)^a
  e^{- b_{\ell} \eta^{(\ell)}} g_\ell(\phi^{(\ell)})\right) = \Theta\left(v- \left(\frac{k_t^{(\ell)}}{Q}\right)^{a+b_\ell}
  \frac{d_\ell g_\ell(\phi^{(\ell)})}{(z^{(\ell)})^{b_\ell}}\right) \notag\\
&
= \Theta\left(v -
  \left(\frac{k_t^{(\ell)}}{Q}\right)^{a+b_\ell}\right) - v\,\delta\left(v -
  \left(\frac{k_t^{(\ell)}}{Q}\right)^{a+b_\ell}\right) \ln
\frac{d_\ell  g_\ell(\phi^{(\ell)})}{(z^{(\ell)})^{b_\ell}} + \dots
\end{align}
where we neglect N$^3$LL corrections in the expansion.

With the above expansion it is sufficient to use the azimuthally
averaged splitting function in $d=4-2\epsilon$ dimensions to construct
the hard-collinear squared matrix element, since in
Eq.~\eqref{eq:theta_expansions} the only term that involves a
non-trivial $\phi^{(\ell)}$ dependence is finite in $d=4$
dimensions. In this approximation, the hard-collinear emission
probability relative to each leg $\ell$ is given by 
\begin{equation}
 \label{eq:M2_hc}
 [dk] M_{\rm hc,\ell}^2(k)=
2 \frac{e^{\gamma_E\epsilon}}{\Gamma(1-\epsilon)} \frac{dk_t^{(\ell)}}{k_t^{(\ell)}} \left(\frac{\mu_R}{k_t^{(\ell)}}\right) ^{2\epsilon}  dz^{(\ell)}\frac{d\phi^{(\ell)}}{2\pi}
\frac{\alpha_s(k_t^{(\ell)})}{2\pi} C_F \left(\frac{1+(1-z^{(\ell)})^2}{z^{(\ell)}} - \epsilon z^{(\ell)}
  - \frac{2}{z^{(\ell)}}\right),
\end{equation}
where the running coupling has been renormalised in the
$\overline{\rm MS}$ scheme. The factor $-2/z^{(\ell)}$ eliminates the
double counting of the soft singularity which is accounted for in the
first line of Eq.~\eqref{eq:cumbasic_NNLL_start}. After performing the
integral in the last line analytically,
Eq.~\eqref{eq:cumbasic_NNLL_start} becomes
\begin{align}\label{eq:cumbasic_NNLL_final}
&\Sigma_{\rm NNLL}(v) = C(\alpha_s(Q)) e^{-R_{\rm s}(v) - R_{\rm hc}(v)}\notag\\
& \times \frac{e^{-R_{\rm s}(\delta v)}}{e^{-R_{\rm s}(v)}}\bigg\{ \sum_{n=0}^\infty \frac{1}{n!} \int\prod_{i=1}^n
  [dk_i] M_{\rm s}^2(k_1,...,k_n) \Theta( v - V(\{\tilde
  p\},k_1,\dots,k_n)) \prod_{\rm clust.}\Theta\left(V_{\rm
  sc}(k_{\rm clust.} )- \delta v\right) \notag\\
& + \sum_{n=0}^\infty \frac{1}{n!}\int\prod_{i=1}^n
  [dk_i] M_{\rm sc}^2(k_i) \Theta\left(V_{\rm
  sc}(k_{i} )- \delta v\right) \notag\\
&\times \sum_{\ell=1}^2\int [dk_{\rm hc}] M_{\rm
  hc,\ell}^2(k_{\rm hc})\bigg(\Theta( v - V(\{\tilde
  p\},k_1,\dots,k_n,k_{\rm hc})) - \Theta( v - V(\{\tilde
  p\},k_1,\dots,k_n))\Theta(v-V_{\rm sc}(k_{\rm hc}))\bigg)\notag\\
&+  \sum_{\ell=1}^2\frac{\alpha_s(Q v^{
    \frac{1}{a+b_\ell}})}{2\pi} C^{(1)}_{\rm hc, \ell}\sum_{n=0}^\infty \frac{1}{n!}\int\prod_{i=1}^n
  [dk_i] M_{\rm sc}^2(k_i) \Theta\left(V_{\rm
  sc}(k_{i} )- \delta v\right)\Theta( v - V(\{\tilde
  p\},k_1,\dots,k_n)) \bigg\},
\end{align} 
where we introduced the hard-collinear radiator defined as
\begin{align}\label{eq:hcrad}
R_{\rm hc}(v) = \sum_{\ell=1}^2\int^{Q^2}_{Q^2 v^{
  \frac{2}{a+b_\ell}}}\frac{d k^2}{k^2}
  \gamma_\ell(\alpha_s(k)).
\end{align}
The hard-collinear constant $C^{(1)}_{\rm hc, \ell}$ is given by 
\begin{equation}
\label{eq:C1-hc}
C^{(1)}_{\rm hc, \ell} = C_F\left(\frac12
  +\frac{b_\ell}{a+b_\ell}\frac{7}{2} + 3 \ln\bar d_\ell  \right) \,
\end{equation}
with
\begin{equation}
\label{eq:lndbar}
\ln\bar d_\ell = \int_0^{2\pi} \frac{d\phi^{(\ell)}}{2\pi}\ln d_\ell g_\ell(\phi^{(\ell)})\,.
\end{equation}
Finally, the constant part of virtual corrections at NNLL is given by 
\begin{equation}
C(\alpha_s(Q))=1+\frac{\alpha_s(Q)}{2\pi} H^{(1)} + {\cal O}(\alpha_s^2),
\end{equation}
where
\begin{equation}
H^{(1)} = C_F \left(\pi^2 - \frac{19}{2}\right)\,.
\end{equation}

Each of the terms in the resolved contribution to
Eq.~\eqref{eq:cumbasic_NNLL_final} can be further decomposed into a
finite set of corrections so that the NNLL cross section
$\Sigma_{\rm NNLL}(v)$ can be parametrised with the following master
formula (we define $\lambda = \alpha_s(Q)\beta_0\ln(1/v)$)
\begin{align}
  \label{eq:Sigma-NNLL}
  \Sigma_{\rm NNLL}(v) = e^{-R_{\rm s}(v) - R_{\rm
      hc}(v)}\Bigg[\FNLL(\lambda)\bigg(1 + \frac{\alpha_s(Q)}{2\pi}
      H^{(1)} &+ \sum_{\ell=1}^2\frac{\alpha_s(Q v^{
    \frac{1}{a+b_\ell}})}{2\pi} C^{(1)}_{\rm hc, \ell}\bigg)\notag\\
&+\frac{\alpha_s(Q)}{\pi}\delta \FNNLL(\lambda)\Bigg]\,,
\end{align}
where the functions $\FNLL$ and $\delta \FNNLL$ have a general
expression for any rIRC safe
observable~\cite{Banfi:2004yd,Banfi:2014sua,Banfi:2016zlc} and can be
efficiently evaluated numerically in $d=4$ dimensions. 
The NNLL function is decomposed as follows
\begin{equation}
  \label{eq:deltaF}
  \delta \FNNLL = \delta \mathcal{F}_{\rm sc}
+\delta
  \mathcal{F}_{\rm hc}+\delta \mathcal{F}_{\rm rec}+\delta
  \mathcal{F}_{\rm wa}+\delta \mathcal{F}_{\rm correl} + \delta
  \mathcal{F}_{\rm clust}\,,
\end{equation}
where each term has a well-defined physical origin.
\begin{itemize}
\item The corrections $\delta \mathcal{F}_{\rm sc}$,
  $\delta \mathcal{F}_{\rm wa}$, $\delta \mathcal{F}_{\rm correl}$ and
  $\delta \mathcal{F}_{\rm clust}$ have soft origin and they all
  originate from the first term in the curly brackets of
  Eq.~\eqref{eq:cumbasic_NNLL_final}. 

  The function $\delta \mathcal{F}_{\rm sc}$ accounts for running
  coupling corrections to the real emissions in the CMW
  scheme~\cite{Catani:1990rr} as well as for the correct rapidity
  boundary for a single soft-collinear emission. For event shapes
  variables this correction is particularly
  simple~\cite{Banfi:2014sua} owing to the fact that the rapidity
  dependence of the observable can be always handled analytically. For
  observables with a more complicated rapidity dependence, such as jet
  rates~\cite{Banfi:2016zlc}, the running coupling correction and the
  correction due to the rapidity boundary must be treated separately.

  The function $\delta \mathcal{F}_{\rm wa}$ accounts for the
  difference between the observable and its soft-collinear
  parametrisation for a single soft-non-collinear (wide-angle)
  emission accompanied by many soft-collinear gluons.

  At NLL all {\it resolved} emissions are strongly ordered in angle,
  and thus emitted independently. The matrix element used to compute
  the function $\FNLL$ is simply given by a product of an arbitrary
  number of single-gluon emission squared amplitudes
  $\tilde M_{\rm sc}^2(k_i)$, in the decomposition of
  Eq.~\eqref{eq:clusters}.
  However, starting from NNLL two or more resolved emissions can
  become close in angle. In this type of configurations, the squared
  amplitude is given by an abelian term (defined by the product of $n$
  single emission probabilities) and by non-abelian, correlated
  clusters of two or more particles (see Eq.~\eqref{eq:clusters}).
  At NNLL it is sufficient to account for the effect of only two
  emissions getting close in angle, while the others can be considered
  far apart. This induces two types of corrections:
  $\delta \mathcal{F}_{\rm correl}$ and $\mathcal{F}_{\rm clust}$.

  The {\it correlated} correction $\delta \mathcal{F}_{\rm correl}$
  accounts for the insertion in the resolved ensemble of
  soft-collinear, independently emitted gluons of a {\it single}
  double-soft cluster $\tilde M_{\rm s,0}^2(k_1,k_2)$ that is defined
  as the non-abelian part of the square of the double-soft
  current~\cite{Banfi:2014sua}. More details on this correction will
  be given in Sec.~\ref{sec:correl}.

  The {\it clustering} correction $\delta\mathcal{F}_{\rm clust}$
  (defined in Ref.~\cite{Banfi:2016zlc}), on the other hand, accounts
  for the contribution of two independently emitted gluons that become
  close in angle. Due to the nature of most event shapes, this
  correction normally vanishes and it becomes different from zero only
  when the observable has a non-trivial dependence on the rapidity of
  the emissions, as for instance in the case of jet
  rates~\cite{Banfi:2016zlc}.

\item The corrections $\delta\mathcal{F}_{\rm hc}$ and $\delta
  \mathcal{F}_{\rm rec}$ originate from the second term in the curly
  brackets of Eq.~\eqref{eq:cumbasic_NNLL_final}, and have a
  hard-collinear nature. The emission of a hard collinear parton
  induces two types of corrections: at the level of the squared
  amplitude (encoded in $\delta\mathcal{F}_{\rm hc}$), and at the level of
  the kinematics, due to the recoil of the whole event against the
  hard-collinear emission (encoded in $\delta
  \mathcal{F}_{\rm rec}$). 
\item Finally, the term 
\begin{equation}
  \label{eq:FNNLL}
  \FNLL(\lambda)\left(1 + \frac{\alpha_s(Q)}{2\pi}
      H^{(1)} + \sum_{\ell=1}^2\frac{\alpha_s(Q v^{
    \frac{1}{a+b_\ell}})}{2\pi} C^{(1)}_{\rm hc, \ell}\right)\,,
\end{equation}
arises from the first and third term in the curly brackets of
Eq.~\eqref{eq:cumbasic_NNLL_final}, where N$^3$LL corrections were
neglected. The function $\FNLL$ is purely NLL~\cite{Banfi:2004yd},
while the multiplying constants $H^{(1)}$ and $C^{(1)}_{\rm hc, \ell}$
induce NNLL corrections.
\end{itemize}
Since the detailed formulation of the functions $\FNLL$ and
$\delta \FNNLL$ is given in
refs.~\cite{Banfi:2004yd,Banfi:2014sua,Banfi:2016zlc}, we do not
report their expressions here, and refer to the original publications.

In the next section we perform the calculation of the Sudakov radiator
at NNLL accuracy. 
Before we proceed, it is important to stress that in the definition of
the unresolved soft radiation given in Eq.~\eqref{eq:resobsbasic} one
clearly has some freedom in deciding precisely how the resolution
variable is defined. In particular, instead of $V_{\rm sc}$, one could
use any observable $V_{\rm res}$ that shares the same leading
logarithms as the full observable $V$ that is being resummed. This is
in fact the only requisite for this method to be applied.
The choice of $V_{\rm sc}$ is mainly due to computational convenience,
as all ingredients in the Sudakov radiator can be computed
analytically for any rIRC safe observable. Choosing another resolution
variable would change the expression in the Sudakov radiator, and
consequently the expression of the functions $\FNLL$ and
$\delta \FNNLL$.

A particularly important aspect of the definition of the resolution
variable concerns the way $V_{\rm sc}$ in Eq.~\eqref{eq:resobsbasic} is
evaluated on the total momentum $k_{\rm clust.}$ of a cluster of more
than one particle. Although the cluster has a non-zero invariant mass,
$V_{\rm sc}$~\eqref{eq:sc-behaviour} does not depend on the mass, and
hence, in the definition resolution scale, the cluster is treated as
if it were {\em massless}.
This will lead to great simplifications in evaluating the Sudakov
radiator, where the integral over the invariant mass can be evaluated
analytically as it will be shown in Sec.~\ref{sec:rad-NNLL}. The
prescription of treating the cluster as massless also impacts the
definition of the correlated correction
$\delta \mathcal{F}_{\rm correl}$, that will be the subject of
Sec.~\ref{sec:correl}.

\section{The Sudakov radiator at NNLL accuracy}
\label{sec:rad-NNLL}

In this section we explicitly compute the Sudakov radiator defined in
Eqs.~(\ref{eq:soft_radiator}) and~(\ref{eq:hcrad}) to NNLL.

\subsection{The soft radiator}

With the choice of resolution variable as in
Eq.~(\ref{eq:sc-behaviour}), the soft radiator reads
\begin{align}\label{eq:softrad}
R_{\text{s}}(v) = \sum_{\ell=1}^2\int \frac{d^4 k}{(2\pi)^4}  w(m^2,k_t^2+m^2)\Theta\left(d_\ell\left(\frac{k_t}{Q}\right)^a e^{-b_\ell \eta^{(\ell)}} g_\ell(\phi^{(\ell)}) -  v\right) \Theta(\eta^{(\ell)})  \ \ .
\end{align}
where the phase space is given in Eq.~(\ref{eq:PSmassive}). Notice
here that the phase-space measure contains the {\em massive} rapidity
$y$ of the web as defined in Eq.~(\ref{eq:PSmassive}), while the
observable is expressed in terms of the rapidity of a massless parton
$\eta^{(\ell)}$. Now we wish to write Eq.~(\ref{eq:softrad}) in a way
in which LL, NLL and NNLL contributions are separated.
The first step to achieve this is to isolate the dependence on
$d_\ell g_\ell(\phi)$ by expanding the observable constraint in
Eq.~(\ref{eq:softrad}) as follows
\begin{align}
&\Theta\left(d_\ell \,g_\ell(\phi^{(\ell)})
  \left(\frac{k_t}{Q}\right)^{a}e^{-b_\ell \eta^{(\ell)}}-v
    \right) \simeq\,
    \Theta\left(\ln \left[\left(\frac{k_t}{Q}\right)^{a}e^{-b_\ell
        \eta^{(\ell)}}\right] -\ln v\right) \notag\\
&  + \delta\left(\ln \left[\left(\frac{k_t}{Q}\right)^{a}e^{-b_\ell \eta^{(\ell)}}\right]-\ln v\right)\ln
   \left( d_\ell \,g_\ell(\phi^{(\ell)})\right)\notag\\
&+\frac{1}{2}\delta'\left(\ln\bigg[\left(\frac{k_t}{Q}\right)^{a}e^{-b_\ell \eta^{(\ell)}}\bigg]-\ln v\right)\ln^2\left(d_\ell \,g_\ell(\phi^{(\ell)})\right)\,,
    \label{eq:exp-theta-soft}
\end{align}
The first term in the above equation starts at LL accuracy, the second
at NLL accuracy, and so on. This gives
\begin{equation}
  \label{eq:Rs-exp}
  R_{\text{s}}(v)\simeq  \sum_\ell\left( R_\ell(v)+R'_\ell(v)\int_0^{2\pi} \frac{d\phi^{(\ell)}}{2\pi} \ln(d_\ell g_\ell(\phi^{(\ell)}))+R''_\ell(v) \int_0^{2\pi} \frac{d\phi^{(\ell)}}{2\pi} \frac{1}{2}\ln^2(d_\ell g_\ell(\phi^{(\ell)}))  \right)\,,
\end{equation}
with
\begin{equation}
  \label{eq:Rell}
  \begin{split}
  R_\ell(v) & = \int \frac{d^4 k}{(2\pi)^4}  w(m^2,k_t^2+m^2)  \Theta\left(\left(\frac{k_t}{Q}\right)^a e^{-b_\ell \eta^{(\ell)}} -  v\right) \Theta(\eta^{(\ell)})\,, \\ \qquad R'_\ell & = -v \frac{dR_\ell(v)}{dv}\,,\qquad 
  R''_\ell = -v \frac{dR'_\ell(v)}{dv}\,.
  \end{split}
\end{equation}
We now concentrate on $R_\ell(v)$. The kinematic boundary for the
rapidity integral is $\ln(Q/\sqrt{k_t^2+m^2})$. Instead of
computing directly the integral in Eq.~\eqref{eq:Rell}, we split it
into the sum of two terms as 
\begin{equation}
  \label{eq:Rell-final}
  R_\ell(v) \simeq R^0_\ell(v) + \delta R_\ell(v)\,,
\end{equation}
where $R^0_\ell(v) $ is defined as in Eq.~\eqref{eq:Rell} but with a
massless rapidity boundary, i.e. $\ln(Q/k_t)$.  This defines a {\em
  massless} radiator, in which $\eta^{(\ell)}$ coincides with $y$, i.e.
\begin{align}
\label{eq:R0-massless}
R^0_\ell(v) = \int \frac{d^4 k}{(2\pi)^4}  w(m^2,k_t^2+m^2) \Theta\left( \ln \frac{Q}{k_t} - \eta^{(\ell)}\right) \Theta\left(\left(\frac{k_t}{Q}\right)^a e^{-b_\ell \eta^{(\ell)}} - v\right) \Theta(\eta^{(\ell)})\,,
\end{align}
that starts at LL accuracy. The function $\delta R_\ell(v)$ defines a
{\em mass correction}, which accounts for the correct rapidity
bound. By inspecting the phase space constraints due to the physical
rapidity bound and the observable, one finds that the rapidity
integral is bounded by $\ln(Q/\sqrt{k_t^2+m^2})$ only when
$k_t > v^{\frac{1}{a+b_\ell}} Q$. This leads to the following
expression for the mass correction to the soft radiator
\begin{align}
\delta R_\ell(v) = \int \frac{d^4 k}{(2\pi)^4}  w(m^2,k_t^2+m^2)
  \Theta(k_t - v^{\frac{1}{a+b_\ell}} Q)\bigg[ & \Theta\left(
  \ln\sqrt{\frac{Q^2}{k_t^2 + m^2}} - \eta^{(\ell)}\right)\notag\\
&- \Theta\left( \ln\sqrt{\frac{Q^2}{k_t^2}} - \eta^{(\ell)}\right) \bigg]\Theta(\eta^{(\ell)})  \ \ . 
\end{align}
The separation of the radiator as in Eq.~\eqref{eq:Rell-final} has a
physical justification.  If one ignores the running of the coupling
constant, then the massless radiator $R^0_\ell(v)$ only contains
double logarithmic terms, while the mass correction $\delta R_\ell(v)$
is purely single logarithmic.

Let us now focus of the massless radiator at
NNLL. Eq.~\eqref{eq:R0-massless} can be easily evaluated by observing
that the resolution variable does not depend on the mass of the
web. Therefore, in Eq.~\eqref{eq:R0-massless} we can freely integrate
over this variable. The integral of the web $w(m^2,k_t^2+m^2)$ over
its invariant mass defines a generalisation of the physical CMW
coupling~\cite{Catani:1990rr}:
\begin{align}
\int_0^\infty \!\! dm^2 \,w(m^2,k_t^2+m^2) \equiv (4\pi)^2\frac{2 C_\ell}{k_t^2} \alpha_s^{\rm phys}(k_t)\,.
\end{align}
The physical coupling $\alpha_s^{\rm phys}$ is related to the $\overline{\rm MS}$ coupling $\alpha_s$ as follows
\begin{align}
\label{eq:CMW_extended}
\alpha_s^{\rm phys} = \alpha_s \left( 1+ \sum_{n=1}^\infty \left(\frac{\alpha_s}{2\pi}\right)^n K^{(n)} \right) \ \ .
\end{align}
The set of constants $K^{(n)}$ is perturbatively calculable and, once
identified, the massless radiator $R_\ell^0(v)$ is fully determined for any
observable, and given by
\begin{align}
\label{eq:R0-massless-CMW}
  R^0_\ell(v) = 2 C_\ell \int\frac{dk_t}{k_t} \frac{\alpha^{\rm phys}(k_t)}{\pi}\int_0^{\ln(Q/k_t)} \!\!\!\!\!\!\! d\eta\,  \Theta\left(\left(\frac{k_t}{Q}\right)^a e^{-b_\ell \eta} - v\right) \,.
\end{align}
At NNLL accuracy, in the expression of $\alpha_s^{\rm phys}$ one needs
to include only $K^{(1)}$ and $K^{(2)}$, whose expressions are
obtained by integrating the web up to order $\alpha_s^3$. This
requires contributions up to the triple-soft current at tree
level~\cite{Berends:1988zn}, the single-soft two loop
current~\cite{Li:2013lsa}, and the double-soft current at one
loop. One obtains
\begin{subequations}
\begin{align}
  \label{eq:Ks}
 K^{(1)} &= C_A
 \left(\frac{67}{18}-\frac{\pi^2}{6}\right)-\frac{5}{9}n_f\,,
 \\
 K^{(2)} &= C_A^2 \left( \frac{245}{24} - \frac{67}{9}\zeta_2
 + \frac{11}{6}\zeta_3 + \frac{11}{5}\zeta_2^2\right) 
+ C_F n_f \left(-\frac{55}{24} + 2\zeta_3\right) \notag\\
& + C_A n_f \left(-\frac{209}{108} + \frac{10}{9}\zeta_2 - \frac{7}{3} \zeta_3\right) 
 - \frac{1}{27} n_f^2 + \frac{\pi\beta_0}{2} \left(C_A\left(\frac{808}{27}-28\zeta_{3}\right)-\frac{224}{54}n_f\right)\,.
\end{align}
\end{subequations}
Note that $K^{(1)}$ is proportional to the two-loop cusp anomalous
dimension, but this is not true any more starting from $K^{(2)}$.
We stress that this is not the case for rIRC unsafe cases, such as
threshold resummation. In this case, one needs to perform the
integration over the web transverse momentum in $d$ dimensions and
subtract the residual collinear singularity in a given factorisation
scheme. The divergent integral over $k_t$ then gives an extra
contribution that enters at same order as $K^{(2)}$ so that the
coefficient of the $\alpha_s^3L^2$ term in the Sudakov coincides with
the cusp anomalous dimension.

For the computation of the mass correction at NNLL, we only need to
consider the web up to $\alpha_s^2$. In fact, the only non-vanishing
contribution arises from the double-emission soft block,
$\tilde{M}_{s,0}^2(k_a,k_b)$, that can be found in
Appendix~\ref{sec:2ps}. Using the rescaled variable
$\mu^2 = m^2/k_t^2$, one finally gets
\begin{align}
\delta R_\ell(v) &= C_\ell \int_{Qv^{\frac{1}{a+b_{\ell}}}}^Q \frac{dk_t}{k_t} \left(\frac{\alpha_s(k_t)}{\pi}\right)^2
 \int_0^\infty \frac{d\mu^2}{\mu^2(1+\mu)} \left(C_A
  \ln\frac{1+\mu^2}{\mu^4} - 2 \pi \beta_0\right) \ln\left(\sqrt{ \frac{1}{1 +
  \mu^2}}\right) \nonumber \\ &
   =\pi \beta_0 \zeta_2  C_\ell \int_{Qv^{\frac{1}{a+b_{\ell}}}}^Q \frac{dk_t}{k_t} \left(\frac{\alpha_s(k_t)}{\pi}\right)^2\,.
\end{align}
From the above derivation we observe that the ${\cal O}(\alpha_s^3)$
correction to the physical coupling~\eqref{eq:CMW_extended} is the
only place where the $\alpha_s^3$ web enters in a NNLL resummation of
any rIRC safe observable. Indeed the resolved real corrections only
involve correlated blocks with up to two soft partons. Therefore, the
physical coupling defined in Eq.~\eqref{eq:CMW_extended} constitutes a
universal ingredient to account for the triple-correlated soft
contribution (and relative virtual corrections due to the
double-correlated soft at one loop as well as single-correlated soft
at two loops) at NNLL accuracy. In particular, it defines a building
block of a parton shower algorithm at this order, that will be
relevant in the context of the current efforts that aim at improving
the accuracy of these
algorithms~\cite{Nagy:2012bt,Martinez:2018ffw,Jadach:2010aa,Hoche:2017hno,Hoche:2017iem,Dulat:2018vuy,Li:2016yez,Dasgupta:2018nvj}.
\subsection{The hard-collinear radiator}

The hard-collinear part of the radiator, defined in
Eq.~(\ref{eq:hcrad}), starts at NLL accuracy. Up to NNLL accuracy, its
expression is
\begin{equation}
  \label{eq:Rhc-NNLL}
  R_{\rm hc} = \sum_{\ell=1}^2 \int_{Q^2 v^{\frac{2}{a+b_\ell}}}^{Q^2} \frac{dk^2}{k^2} \frac{\alpha_s(k)}{2\pi} \left[\gamma^{(0)}_\ell+\left(\frac{\alpha_s}{2\pi}\right) \gamma^{(1)}_\ell\right]\,.
\end{equation}
In our case, the coefficients $\gamma^{(0)}_\ell$ and
$\gamma^{(1)}_\ell$ are the coefficients of the $\delta(1-x)$ piece of
the $P_{\text{qq}}(x)$ splitting functions with an overall minus sign. In particular, $\gamma^{(0)}_\ell$ is given in Eq.~(\ref{eq:gamma0}), and
\begin{align}
\gamma^{(1)}_\ell &= - \frac{C_F}{2} \left(C_F \left(\frac34 - \pi^2 + 12 \zeta_3\right) + C_A \left( \frac{17}{12} + \frac{11 \pi^2}{9} - 6 \zeta_3 \right) - n_f \left(\frac{1}{6} + \frac{2\pi^2}{9}\right) \right)\,.
\end{align}

\subsection{The radiator up to NNLL accuracy}
The computation of the radiator proceeds by integrating the equations
of the above sections with a running coupling. In particular, at NNLL
accuracy, this is given by the renormalisation group equation
\begin{align}\label{eq:RG}
\mu_R^2 \frac{d\alpha_s}{d \mu_R^2} = -\beta_0 \alpha_s^2 - \beta_1 \alpha_s^3 - \beta_2 \alpha_s^4,
\end{align} 
where $\beta_0$ is given in Eq.~(\ref{eq:beta0}), and the other
coefficients of the beta function, in the $\overline{\text{MS}}$
scheme, are given by
\begin{align}\label{betacoeffs}
\nonumber
\beta_1 &= \frac{17 C_A^2 - 5 C_A n_f - 3 C_F n_f}{24 \pi^2}\,, \\
\beta_2 &= \frac{2857 C_A^3 + (54 C_F^2 - 615 C_F C_A - 1415 C_A^2)n_f + (66 C_F + 79 C_A) n_f^2}{3456 \pi^3} \ \ .
\end{align}
For resummation purposes, it suffices to solve Eq.~\eqref{eq:RG} using
the following ansatz
\begin{align}
\alpha(\mu_R) = \sum_{n=1}^{\infty} \left(\frac{\alpha_s(Q)}{1 + t } \right)^n f_n(t)\,,  \qquad t \equiv \alpha_s \beta_0 \ln(\mu_R^2/Q^2) \ \ .
\end{align}
Plugging the above in Eq.~(\ref{eq:RG}), one finds
\begin{align}
& f_1(t) = 1\,, \quad f_2(t) = - \frac{\beta_1}{\beta_0}\ln(1 + t)\,, \\ & f_3(t) = -\frac{\beta_2}{\beta_0}t +\left(\frac{\beta_1}{\beta_0}\right)^2 [t + (\ln (1+t) - 1)]\ln(1+t)\,,
\end{align}
and we can neglect the contributions of $f_4(t)$ and beyond, since
they start to matter from N$^3$LL accuracy.

It is customary to express the radiator in terms of
$\lambda=\alpha_s \beta_0 \ln(1/v)$, with $\alpha_s=\alpha_s(Q)$. We
further parametrize the radiator in terms of functions of $\lambda$,
in such a way as to separate LL, NLL and NNLL contributions:
\begin{align}
  R_{\text{s}}(v)&=\sum_\ell\left( R_\ell(v)+R'_\ell(v)\int_0^{2\pi}
    \frac{d\phi^{(\ell)}}{2\pi} \ln(d_\ell g_\ell(\phi^{(\ell)}))+R''_\ell(v)
    \int_0^{2\pi} \frac{d\phi^{(\ell)}}{2\pi} \frac{1}{2}\ln^2(d_\ell
    g_\ell(\phi^{(\ell)}))  \right)\,,\\
R_\ell(v) &\simeq R^0_\ell(v) + \delta R_\ell(v)\,,\\
R^0_\ell(v) &= - \frac{\lambda}{\alpha_s \beta_0} g_1^{(\ell)}(\lambda) - g_2^{(\ell)}(\lambda) - \frac{\alpha_s}{\pi} g_3^{(\ell)}(\lambda)\,, \\
\delta R_\ell(v) &=- \frac{\alpha_s}{\pi} \delta g_3^{(\ell)}(\lambda)\,, \\
R_{\text{hc},\ell}(v)  &= - h_2^{(\ell)}(\lambda) - \frac{\alpha_s}{\pi} h_3^{(\ell)}(\lambda)\,, 
\end{align}
where 
\begin{align}
  g_1^{(\ell)} (\lambda) &= \frac{C_\ell}{2} \frac{ (a+b_\ell-2\lambda) \ln \left(1 - \frac{2\lambda}{a+b_\ell}\right) - (a - 2 \lambda ) \ln \left(1 - \frac{ 2 \lambda }{a}\right)}{\pi  b_\ell \beta _0 \lambda}\,, \\\nonumber
  g_2^{(\ell)} (\lambda) &= \frac{C_\ell}{2} \bigg[ \frac{ K^{(1)} \left(a \ln \left( 1 - \frac{2 \lambda
                           }{a}  \right)- (a+b_\ell) \ln \left( 1 - \frac{2 \lambda
                           }{a+b_\ell}\right)\right)}{2 \pi ^2 b_\ell \beta _0^2}  \\\nonumber
                         &+  \frac{\beta _1 (a+b_\ell) \ln ^2\left( 1 -\frac{2\lambda}{a+b_\ell}  \right)}{ 2 \pi b_\ell \beta_0^3} + \frac{\beta_1 (a+b_\ell)
                           \ln \left(1 - \frac{2 \lambda }{a+b_\ell}\right)}{\pi b_\ell \beta_0^3} \\
                         & - \beta_1\frac{a \ln \left( 1 - \frac{2 \lambda }{a}\right) \left(\ln \left(1- \frac{2 \lambda
                           }{a}\right)+2\right)}{2 \pi b_\ell \beta_0^3} \bigg]\,, \\\nonumber
\end{align}
\begin{align}
\nonumber
  g_3^{(\ell)} (\lambda) &= \frac{C_\ell}{2} \bigg[ K^{(1)}  \frac{
                           \beta _1 \left( a^2 (a+b_\ell+2 \lambda )
                           \ln \left( 1 - \frac{2 \lambda }{a}\right)
                           - (a+b_\ell)^2 (a-2\lambda) \ln \left( 1 -
                           \frac{2 \lambda }{a+b_\ell}\right) + 6
                           b_\ell \lambda^2 \right)}{2 \pi b_\ell
                           \beta_0^3 (a-2\lambda) (a+b_\ell-2\lambda)}
  \\\nonumber
&+\frac{\left(\text{$\beta_1 $}^2 (a+b_\ell)^2 (a-2 \lambda ) \ln ^2\left(1-\frac{2 \lambda }{a+b_\ell}\right)-4 b_\ell \lambda ^2 \left(\text{$\beta_0 $}
   \text{$\beta_2 $}+\text{$\beta_1 $}^2\right)\right)}{2 b_\ell
\text{$\beta_0 $}^4 (a-2 \lambda ) (a+b_\ell-2 \lambda )}\\\nonumber
&-\frac{a \ln \left(1-\frac{2 \lambda }{a}\right) \left(2 \text{$\beta_0 $} \text{$\beta_2 $} (a-2 \lambda )+a \text{$\beta_1 $}^2 \ln
   \left(1-\frac{2 \lambda }{a}\right)+4 \text{$\beta_1 $}^2 \lambda
 \right)}{2 b_\ell \text{$\beta_0 $}^4 (a-2 \lambda )}\\\nonumber
&+\frac{(a+b_\ell) \ln \left(1-\frac{2 \lambda }{a+b_\ell}\right) \left(\text{$\beta_0 $} \text{$\beta_2 $} (a+b_\ell-2 \lambda )+2 \text{$\beta_1 $}^2 \lambda
   \right)}{b_\ell \text{$\beta_0 $}^4 (a+b_\ell-2 \lambda )}\\
                         &- K^{(2)} \frac{ 2 \lambda^2 }{4\pi^2 (a-2\lambda) (a+ b_\ell - 2\lambda) \beta_0^2} \bigg]\,,\\
  \delta g_3^{(\ell)}(\lambda) &= -C_\ell\zeta_2 \frac{\lambda }{(a+b_\ell-2\lambda)}\,,\\
  h_2^{(\ell)}(\lambda) &= \frac{\gamma^{(0)}_\ell}{2 \pi \beta_0} \ln\left(1-\frac{2\lambda}{a+b_\ell}\right)\,,  \\
  h_3^{(\ell)}(\lambda) &= \gamma^{(0)}_\ell\frac{\beta_1 \left((a+b_\ell) \left(\ln \left( 1- \frac{2 \lambda
                          }{a+b_\ell}\right)\right)+ 2\lambda \right)}{ 2\beta^2_0 \left(a + b_\ell - 2 \lambda\right)} -\gamma^{(1)}_\ell \frac{\lambda}{2\pi \beta_0 (a+b_\ell-2\lambda)} \ \ .
\end{align}
The corresponding results for $b_\ell=0$ can be obtained by taking the
limit of the above expressions for $b_\ell \to 0$. We also include
here various derivatives of the {\em massless} radiator that appear in
the evaluation of the correction functions: 
\begin{align}
R^\prime_{\text{NLL},\ell} &= \frac{C_\ell}{b_{\ell} \pi \beta_0} \left( \ln \left( 1- \frac{2\lambda}{a+b_{\ell}} \right) -  \ln \left( 1- \frac{2\lambda}{a} \right) \right)\,, \\ \nonumber
R^\prime_{\text{NNLL},\ell} &= \frac{C_\ell \alpha_s(Q)}{b_{\ell} \pi^2 \beta_0^2 (a-2\lambda) (a+b_{\ell} -2\lambda)} \bigg[ b_{\ell} \beta_0 \lambda K^{(1)} - 2 \pi b_{\ell} \beta_1 \lambda \\
&- \pi a (a+b_{\ell} -2 \lambda) \beta_1 \ln \left( 1- \frac{2\lambda}{a} \right) + \pi (a+b_{\ell}) (a-2\lambda) \beta_1 \ln \left( 1- \frac{2\lambda}{a+b_{\ell}} \right) \bigg]\,, \\
R^{\prime\prime}_{\ell} &\equiv \alpha_s(Q) \beta_0 \frac{d R^\prime_{\text{NLL},\ell}}{d \lambda} =  2 C_\ell \frac{\alpha_s(Q)}{\pi} \frac{1}{(a-2\lambda)(a+b_{\ell}-2\lambda)}\,.
\end{align}

\subsection{The correlated correction $\mathbf{\dF{correl}}$}
\label{sec:correl}

The definition of the correlated correction
$\delta \mathcal{F}_{\rm correl}$ is given
by~\cite{Banfi:2014sua}\footnote{We included the multiplicity factor
  $1/2!$ although the term $\tilde M_{s,0}^2(k_a,k_b)$ also contains
  the contribution of two quarks. The corresponding term in the
  squared amplitude is then multiplied by two. The final expression is
reported in Appendix~\ref{sec:2ps}.}
\begin{equation} 
\label{eq:dF-correl} 
\begin{split}
  &\frac{\alpha_s(Q)}{\pi}\dF{correl}(\lambda)=e^{-\int_{\delta v}^{v}[dk] M^2_{\rm
      sc}(k)}\sum_{n=0}^{\infty}\frac{1}{n!}\int_{\delta v}
  \prod_{i=1}^n [dk_i] M^2_{\rm sc}(k_i) \frac{1}{2!}\int[dk_a] [dk_b]
  \tilde M_{s,0}^2(k_a,k_b) \times \\ \times
  &\left[\Theta\left(v-\Vsc{k_a,k_b,k_1,\dots,k_n}\right)-\Theta\left(v-
      \lim_{m^2\to0}\Vsc{k_a+k_b,k_1,\dots,k_n}\right)\right]\,,
\end{split} 
\end{equation}
where $k_a$ and $k_b$ are the two soft emissions close in angle and
collinear to the same leg, while the remaining soft-collinear
emissions $k_i$ have very disparate angles, and hence are emitted
independently. The invariant mass of the web is denoted by
$m^2=(k_a+k_b)^2$. The configuration in which $k_a$ and $k_b$ are
collinear to different Born legs requires the parent gluon to be
emitted with a large angle, and hence gives at most a N$^3$LL
contribution. The observable
\begin{equation}
\Vsc{k_1,\dots,k_n}
\end{equation}
simply denotes the soft-collinear approximation of the full
observable, as this is the only limit relevant here.

In order to make sense of the difference between the two $\Theta$
functions, we should first discuss how the cluster of two soft partons
is treated in the {\tt ARES} algorithm.

The two-particle correlated soft block $\tilde M_{\rm s,0}^2(k_a,k_b)$
diverges when $k_a$ and $k_b$ are collinear, i.e. when the invariant
mass of the cluster tends to zero. Such a divergence is entirely
cancelled by the one-loop correction to the single-emission cluster
$\tilde M_{\rm s,1}^2(k)$, that should be included at NNLL. In the
evaluation of the Sudakov radiator, as shown in
Sec.~\ref{sec:rad-NNLL}, the cancellation of the above collinear
singularity is performed analytically. This is because we chose to
define a resolution variable that does not depend on the invariant
mass of the cluster, and hence the corresponding integral becomes
straightforward. This inclusive integration of
$\tilde M_{\rm s,0}^2(k_a,k_b)$ and $\tilde M_{\rm s,1}^2(k)$ is
responsible for the presence of the physical coupling in the radiator,
see Eq.~\eqref{eq:CMW_extended}.

In the resolved radiation, however, the situation is more complicated,
as in the full observable we are not allowed to integrate over the
invariant mass of the cluster inclusively. We can, however, define the
following subtraction scheme to cancel the collinear singularity
arising from $\tilde M_{\rm s,0}^2(k_a,k_b)$ in four dimensions. We
first treat the $\{k_a,k_b\}$ cluster inclusively, as done in the
definition of the radiator. This can be done by considering only the
total momentum $k_a+k_b$ when evaluating the contribution of the
cluster to the observable, and treat it as if it were a massless
(lightlike) momentum in the computation of the observable.
This once again allows us to combine it with the one loop correction
to the single-emission cluster $\tilde M_{\rm s,1}^2(k)$
analytically. This contribution is encoded in
$\delta \mathcal{F}_{\rm sc}$ which features the coefficient $K^{(1)}$
of Eq.~(\ref{eq:Ks}), as explained in
refs.~\cite{Banfi:2014sua,Banfi:2016zlc}. As a second step, we
consider the difference between the full observable, where the
$\{k_a,k_b\}$ cluster is treated exclusively, and its inclusive
approximation that we considered above to cancel the singularity
against the virtual correction.  This is represented by the difference
in the two $\Theta$ functions in Eq.~\eqref{eq:dF-correl}. It is
important to bear in mind that in the second $\Theta$ function the
observable $V_{\rm sc}$ treats the momentum $k_a+k_b$ as if it were
{\em massless}, in order to exactly match our convention for the
cancellation of real and virtual corrections in
$\delta \mathcal{F}_{\rm sc}$. This is implemented by the limit in the
second $\Theta$ function in Eq.~\eqref{eq:dF-correl}.

There are many ways to parametrise the phase space for $k_a$ and $k_b$
to keep, in $\dF{correl}$, only NNLL contributions that correspond to
configurations in which $k_a$ and $k_b$ are collinear to the same Born
leg. One possible parametrisation was presented in
Ref.~\cite{Banfi:2014sua}. Instead, here we adopt the notation of
appendix~\ref{sec:2ps}, that is the same we used to compute the NNLL
radiator. 
We define the rescaled invariant mass of the $\{k_a,k_b\}$ cluster
$\mu^2=m^2/k_t^2$, and we introduce the {\it pseudo-parent} parton $k$
of $k_a$ and $k_b$ with transverse momentum $k_t$ and observable
fraction $\zeta$ defined as
\begin{equation}
  \label{eq:k-def}
  \vec k_t = \vec k_{t,a}+\vec k_{t,b}\,,\qquad\zeta \equiv \frac{V_{\rm sc}(k_a+k_b)}{v}\,.
\end{equation}
Using Eqs.~(\ref{eq:ps2-take3}) and~(\ref{eq:tildeM2}), the squared
amplitude for a double-soft correlated emission reads
\begin{equation}
  \label{eq:Mtilde-int}
  \frac{1}{2!}[dk_a] [dk_b] \tilde M_{s,0}^2(k_a,k_b) = [dk] M^2_{\rm sc}(k) \frac{\alpha_s(k_t)}{2\pi} \frac{d\mu^2}{\mu^2(1+\mu^2)} dz \frac{d\phi}{2\pi}\frac{1}{2!}C_{ab}(\mu,z,\phi)\,,
\end{equation}
with $C_{ab}(\mu,z,\phi)=C(k_a,k_b)$ given by Eq.~(\ref{eq:Corr}), and
$\mu^2\in [0,\infty)$, $z\in[0,1]$, $\phi\in[0,2\pi)$. The matrix
element squared and phase space for the pseudo-parent $k$ is given in
Eq.~(\ref{eq:matrix-element}).  Actually, due to the fact that the
pseudo-parent has a non-zero invariant-mass, the integral over its
rapidity $\eta^{(\ell)}$ should have the boundary
$|\eta^{(\ell)}|<\ln(Q/\sqrt{k_t^2+m^2})$. However, following what is
done in the computation of the NLL function
$\FNLL$~\cite{Banfi:2004yd}, we observe that the exact position of the
rapidity integration bound in the resolved radiation enters at one
logarithmic order higher. Therefore, in order to neglect all N$^3$LL
corrections and obtain a result that is purely NNLL, we replace the
actual rapidity integration limit with the massless one, as done in
Eq.~(\ref{eq:matrix-element}).
The integral over $\eta^{(\ell)}$ can be evaluated
analytically~\cite{Banfi:2014sua} and the correlated correction takes
the following simple form
\begin{align}
& \dF{correl}(\lambda)  =
    \int_0^\infty\frac{d\zeta}{\zeta} \sum_{\ell=1,2} \int_0^{2\pi} \frac{d
      \phi^{(\ell)}}{2\pi}\left(\frac{\lambda}{2a
        \beta_0}
      \frac{R''_\ell(v)}{\alpha_s(Q)}\right)\int_0^\infty\frac{d\mu^2}{\mu^2(1+\mu^2)}
    \int_0^1 dz
    \int_0^{2\pi}\frac{d\phi}{2\pi}\frac{1}{2!}C_{ab}(\mu,z,\phi)\times\notag
    \\ & \times \epsilon^{R'_{\mathrm{NLL}}}
   \sum_{n=0}^{\infty}\frac{1}{n!} \prod_{i=1}^n
    \int_{\epsilon}^{\infty} \frac{d\zeta_i}{\zeta_i}\sum_{\ell_i=1,2} \int_0^{2\pi}
   \frac{d\phi_i^{(\ell)}}{2\pi} R'_{\mathrm{NLL},
     \ell_i}\times \notag\\
  \label{eq:Fcorrel-new}
&\times \,\left[ \Theta\left(1-\lim_{v\to 0}\frac{\Vsc{k_a,k_b,k_1,\dots,k_n}}{v}\right)-\Theta\left(1-\lim_{\mu^2\to 0}\lim_{v\to 0}\frac{\Vsc{k_a+k_b,k_1,\dots,k_n}}{v}\right) \right]\,.
\end{align}

Naturally, for analytic calculations the parametrisation for the phase
space and matrix element should be chosen in order to simplify the
integrals for any given observable. The present choice will make the
integrations in the next section simpler, while an alternative
parametrisation was reported in Ref.~\cite{Banfi:2014sua}. 
\subsubsection{Additive observables}
\label{sec:Fcorrel-additive}
 A particularly interesting case is
that of additive observable, for which all NNLL corrections admit a
simple analytic form, reported in the appendix of
Ref.~\cite{Banfi:2014sua}. The correlated correction for such
observables can be simplified considerably. The additivity of the
observable implies that
\begin{equation}
  \label{eq:Vcorr-additive}
  \begin{split}
  \Vsc{k_a,k_b,k_1,\dots,k_n}& =V_{\rm sc}(k_a)+V_{\rm
    sc}(k_b)+\Vsc{k_1,\dots,k_n}\,,\\
\Vsc{k_a+k_b,k_1,\dots,k_n}& =V_{\rm sc}(k_a+k_b)+\Vsc{k_1,\dots,k_n}\,.
  \end{split}
\end{equation}
We now introduce $f_{\rm correl}^{(\ell)}(\mu,z,\phi)$ as:
\begin{equation}
  \label{eq:fcorrel-def}
  V_{\rm sc}(k_a)+V_{\rm
    sc}(k_b)= \zeta v\, f^{(\ell)}_{\rm correl}(\mu,z,\phi)\,. 
\end{equation}
This gives 
\begin{equation}
  \label{eq:Vcorr-additive-zeta}
  \begin{split}
  \lim_{v\to 0}\frac{\Vsc{k_a,k_b,k_1,\dots,k_n}}{v}& =\zeta\, f^{(\ell)}_{\rm correl}(\mu,z,\phi)+\lim_{v\to 0}\frac{\Vsc{k_1,\dots,k_n}}{v} \,,\\
\lim_{\mu^2\to 0}\lim_{v\to 0}\frac{\Vsc{k_a+k_b,k_1,\dots,k_n}}{v}& =\zeta +\lim_{v\to 0}\frac{\Vsc{k_1,\dots,k_n}}{v} \,.
  \end{split}
\end{equation}
Following the derivation in the appendix of Ref.~\cite{Banfi:2014sua},
we can now rescale the momenta $k_1,\dots,k_n$ in two ways. In Eq.~(\ref{eq:Fcorrel-new}), in the term involving the step function of $\Vsc{k_a,k_b, k_1,\dots, k_n}$, we construct rescaled momenta $\tilde k_1,\dots,\tilde k_n$ such that
\begin{equation}
  \label{eq:Vab-rescaled}
\Vsc{\tilde k_1,\dots,\tilde k_n}= \frac{\Vsc{k_1,\dots,k_n}}{1-\zeta
f^{(\ell)}_{\rm correl}}\,.
\end{equation}
In the term with the step function of $\Vsc{k_a+k_b, k_1,\dots, k_n}$
we construct another set of rescaled momenta
$\tilde k'_1,\dots,\tilde k'_n$ such that
\begin{equation}
  \label{eq:Va+b-rescaled}
   \Vsc{\tilde k_1',\dots,\tilde k_n'}= \frac{\Vsc{k_1,\dots,k_n}}{1-\zeta}\,. 
\end{equation}
Performing similar formal manipulations as in
Ref.~\cite{Banfi:2014sua}, we obtain
\begin{equation}
  \label{eq:Fcorrel-additive-take2}
  \begin{split}
    \dF{correl}(\lambda) &=
    \int_0^\infty\frac{d\zeta}{\zeta}\sum_{\ell=1,2}\int_0^{2\pi} \frac{d
      \phi^{(\ell)}}{2\pi} \left(\frac{\lambda}{2a
        \beta_0}
      \frac{R''_\ell(v)}{\alpha_s(Q)}\right)\int_0^\infty\frac{d\mu^2}{\mu^2(1+\mu^2)}
    \int_0^1 dz
    \int_0^{2\pi}\frac{d\phi}{2\pi}\frac{1}{2!}C_{ab}(\mu,z, \phi)\times
    \\ \times &  \left\{  \left(1-\zeta f^{(\ell)}_{\rm
          correl}(\mu,z,\phi)\right)^{R'}\Theta\left(1-\zeta f^{(\ell)}_{\rm
          correl}(\mu,z,\phi)\right) \right.\notag\\
&\left.\times \int d{\cal Z}[\{R'_{\mathrm{NLL}, \ell_i}, \tilde k_i\}] \,\Theta\left(1-\lim_{v\to 0}\frac{\Vsc{\tilde k_1,\dots,\tilde k_n}}{v}\right)\right. \\ & \left.-\left(1-\zeta \right)^{R'}\Theta\left(1-\zeta \right)\int d{\cal Z}[\{R'_{\mathrm{NLL}, \ell_i}, \tilde k'_i\}]\Theta\left(1-\lim_{v\to 0}\frac{\Vsc{\tilde k'_1,\dots,\tilde k'_n}}{v}\right) \right\}\,.
  \end{split}
\end{equation}
After performing the $\zeta$ and $\phi^{(\ell)}$ integrations analytically, we
obtain the factorised form
\begin{equation}
  \label{eq:Fcorrel-additive-final}
  \begin{split}
    \dF{correl}(\lambda) & =
     -\FNLL(\lambda)\sum_{\ell=1,2}\left(\frac{\lambda}{2a
        \beta_0}
      \frac{R''_\ell(v)}{\alpha_s(Q)}\right)\times \\ & \times \int_0^\infty\frac{d\mu^2}{\mu^2(1+\mu^2)}
    \int_0^1 dz
    \int_0^{2\pi}\frac{d\phi}{2\pi}\frac{1}{2!}C_{ab}(\mu,z,\phi)\ln f^{(\ell)}_{\rm correl}(\mu,z,\phi)\,,
  \end{split}
\end{equation}
in which the correlated correction reduces to a number that multiplies
the NLL function $\FNLL$. This result will be used in
Sec.~\ref{sec:FC} to obtain the NNLL resummation for angularities and
moments of energy-energy correlation.

\section{Fully worked out examples: angularities and moments of EEC}
\label{sec:FC}

In this section we apply the resummation procedure described in the
previous sections to interesting observables in $e^+e^-$ annihilation,
namely angularities and moments of energy-energy correlation.

Angularities are defined with respect to some reference axis, usually
the thrust~\cite{Berger:2003iw} or the winner-take-all
(WTA)~\cite{Larkoski:2013eya,Larkoski:2014uqa} axis. They depend on a
parameter $x$, as follows
\begin{equation}
  \label{eq:taux}
  \tau_x \equiv \frac{\sum_i E_i |\sin\theta_i|^x(1-|\cos\theta_i|)^{1-x}}{\sum_i |\vec q_i|}\,,
\end{equation}
where the sum runs over all hadrons in the event, $(E_i,\vec q_i)$ is
the four-momentum of hadron $i$, and $\theta_i$ is the angle between
hadron $i$ and the reference axis.

In Ref.~\cite{Banfi:2004yd}, another class of observables was
introduced, the fractional moments of energy-energy correlation (EEC),
defined by
\begin{equation}
  \label{eq:FC-def}
  FC_x = \sum_{i\ne j} \frac{E_i E_j |\sin\theta_{ij}|^x(1-|\cos\theta_{ij}|)^{1-x}}{\left(\sum_i E_i\right)^2}\Theta\left[(\vec q_i\cdot n_T)(\vec q_j\cdot \vec n_T)\right]\,,
\end{equation}
where, as before, the sums run over all hadrons in the event,
$\theta_{ij}$ denotes the angle between hadrons $i$ and $j$, and
$\vec n_T$ is the thrust axis. Note that similar variables have
attracted interest due to their discriminating power between quark-
and gluon-initiated jets~\cite{Larkoski:2013eya}. For instance, in jet
studies for $e^+e^-$ collisions, one considers
\begin{equation}
  \label{eq:Cbeta}
  C_1^{(\beta)} \equiv \sum_{i\ne j} \frac{E_i E_j}{Q^2} \theta_{ij}^\beta\,,
\end{equation}
where one considers the particles $i,j$ within a given jet.  At hadron
colliders the definition of $C_1^{(\beta)}$~\cite{Larkoski:2013eya}
involves the transverse momentum and the angular distance
$R_{ij}^2 = \Delta y_{ij}^2 + \Delta\phi_{ij}^2$ between final state
particles.
The global component of the resummed cross section for these
observables has analogous resummation properties as the observables
$FC_{2-\beta}$ studied here. However, in this case, the cross section
receives a non-global logarithmic correction starting at NLL.
Both angularities with respect to the WTA axis and moments of EEC have
the property that, in the presence of multiple soft and collinear
emissions $k_1,\dots,k_n$, they are always additive, i.e.
\begin{equation}
  \label{eq:Vsc-additive}
  V_{\rm sc}(\{\tilde p\},k_1,\dots,k_n) = \sum_{i=1}^n V_{\rm sc}(k_i)\,. 
\end{equation}
Angularities with respect to the thrust axis are additive as long as
$x<1$~\cite{Banfi:2004yd}. For these observables, $x<1$ is the range of
values of $x$ that we will implicitly consider in the following. 

The NNLL resummed distribution is given by Eq.~(\ref{eq:Sigma-NNLL}). Our
task is to compute each ingredient of that formula.
First, we observe that both $\tau_x$ and $FC_x$ are infrared safe, and
collinear safe for $x<2$. With a single soft and collinear emission,
and in the range of values of $x$ appropriate for each observable, we
have
\begin{equation}
  \label{eq:Vsc-k}
  V_{\rm sc}(k) = \frac{k_t}{Q}e^{-(1-x) \eta^{(\ell)}}\,,\qquad \ell=1,2\,.
\end{equation}
Therefore, the soft-collinear radiator $R_{\rm s}(v)$, the
hard-collinear radiator $R_{\rm hc}(v)$ and the hard-collinear
constant $C^{(1)}_{{\rm hc},\ell}$ are obtained by computing
Eqs.~(\ref{eq:Rs-exp}),~(\ref{eq:Rhc-NNLL}) and~\eqref{eq:C1-hc}
respectively, with $a=1$, $b_\ell=1\!-\! x$, and
$d_\ell=g_\ell(\phi)=1$. In the following subsections, we compute the
corrections due to real radiation.

The results we will present below for the WTA-axis angularities have
been found to be in complete agreement with the findings of
Ref.~\cite{Procura:2018zpn}, that have been obtained in a SCET
framework.

\subsection{Soft-collinear corrections}
\label{sec:FC-sc-corr}
Since the observables we consider are additive, they fall into the
category studied in appendix C of Ref.~\cite{Banfi:2014sua}.
This gives
\begin{align}
  \label{eq:Fnll}
    \cF_\NLL(\lambda) & = \frac{e^{-\gamma_E
        R'_\NLL}}{\Gamma(1+R'_\NLL)}\,,\\ \dF{sc}(\lambda)&=
    -\frac{\pi}{\alpha_s(Q)}\cF_\NLL(\lambda) \bigg[\delta
      R'_\NNLL\left(\psi^{(0)}(1+R'_\NLL)+\gamma_E\right)\notag\\
  \label{eq:Fsc}
&+\frac{R''}{2}\left(
        \left(\psi^{(0)}(1+R'_\NLL)+\gamma_E\right)^2-\psi^{(1)}(1+R'_\NLL)+\frac{\pi^2}{6}
      \right)\bigg]\,.
\end{align}
Here, for notational convenience, we introduced
$R'_\NLL = R'_{\NLL,1}+R'_{\NLL,2}$, and similarly for $R'_\NNLL$ and
$R''$.  Note that both expressions in Eqs.~\eqref{eq:Fnll}
and~\eqref{eq:Fsc} do not depend explicitly on the parameter $x$, but
this dependence is implicit in the functions $R'_\NLL$, $R'_\NNLL$ and
$R''$.

\subsection{Hard-collinear and recoil corrections}
\label{sec:FC-hc-corr}

If we add to an ensemble of soft and collinear emissions
$k_1,\dots,k_n$ a single hard emission $k$, collinear to
either $p_1$ or $p_2$, our observables behave as follows
\begin{equation}
  \label{eq:FC-hc}
  V_{\rm hc}(\ptilde,k,k_1,\dots,k_n) - V(\ptilde,k_1,\dots,k_n) = \left(\frac{k_t}{Q}\right)^{2-x} f_{\rm hc}^{(\ell)}(z,\phi^{(\ell)})\,.
\end{equation}
For the angularities with respect to the thrust axis, we obtain
\begin{equation}
  \label{eq:fhc-taux-thrust}
  f_{\rm hc}^{(\ell)}(z,\phi^{(\ell)})= \frac{z^{1-x}+(1-z)^{1-x}}{[z(1-z)]^{1-x}}\,,\qquad x<1\,,
\end{equation}
whereas if we compute angularities with respect to the WTA axis, we
obtain
\begin{equation}
  \label{eq:fch-taux-WTA}
  f_{\rm hc}^{(\ell)}(z,\phi^{(\ell)})= \frac{1}{[z(1-z)]^{1-x}}
\frac{1}{\max[z,1-z]}\,,\qquad x<2\,.
\end{equation}
Finally, for fractional moments of EEC, we get
\begin{equation}
  \label{eq:fhc-FCx}
  f_{\rm hc}^{(\ell)}(z,\phi^{(\ell)}) = \frac{1}{[z(1-z)]^{1-x}}\,,\qquad x<2 \,.
\end{equation}
If we extrapolate $f_{\rm hc}^{(\ell)}(z,\phi^{(\ell)})$ for $z\to 0$
we obtain the same result for all observables:
\begin{equation}
  \label{eq:FC-sc-extrap}
f_{\rm hc}^{(\ell)}(z,\phi^{(\ell)}) \to \frac{1}{z^{1-x}} \equiv f_{\rm sc}^{(\ell)}(z,\phi^{(\ell)}) \,.
\end{equation}
The function $f_{\rm sc}^{(\ell)}$ is the only one needed to compute
the correction $\delta\mathcal{F}_{\rm hc}$ according to the procedure
described in appendix C of Ref.~\cite{Banfi:2014sua}, which leads to
\begin{equation}
  \label{eq:dF-hc}
  \begin{split}
    \dF{hc}& =
    \left(\psi^{(0)}(1+R'_\NLL)+\gamma_E\right)\cF_\NLL(\lambda)\times 
    \frac{3}{2} C_F \sum_{\ell=1}^2 \frac{\alpha_s(v^{1/(a+b_\ell)}
      Q)}{\alpha_s(Q)(a+b_\ell)}\\ & = \frac{3 C_F}{2-x}\frac{\alpha_s(v^{1/(2-x)}
      Q)}{
      \alpha_s(Q)}\left(\psi^{(0)}(1+R'_\NLL)+\gamma_E\right)\cF_\NLL(\lambda)\,.
  \end{split}
\end{equation}
For the recoil correction $\dF{rec}$ we need both
$f_{\rm hc}^{(\ell)}$ and $f_{\rm sc}^{(\ell)}$. Specialising the
formulae of appendix C of Ref.~\cite{Banfi:2014sua} to the present
case, for the angularities with respect to the thrust axis, we obtain
\begin{equation}
\label{eq:Frec-taux-thrust}
\begin{split}
   & \dF{rec}  = \cF_\NLL(\lambda) \sum_{\ell=1}^2 \frac{\alpha_s(v^{1/(a+b_\ell)}
    Q)}{\alpha_s(Q)(a+b_\ell)} \int_0^1\!\! dz\, C_F \frac{1+(1-z)^2}{z}\int_0^{2\pi}\frac{d\phi^{(\ell)}}{2\pi} \ln\frac{f_{\rm sc}^{(\ell)}(z,\phi^{(\ell)})}{f_{\rm hc}^{(\ell)}(z,\phi^{(\ell)})} \\ & 
=\frac{2 C_F}{2-x} \frac{\alpha_s(v^{1/(2-x)} Q)}{\alpha_s(Q)}\cF_\NLL(\lambda) \int_0^1\!\! dz \frac{1+(1-z)^2}{z}\left[\ln((1-z)^{1-x})-\ln\left(z^{1-x}+(1-z)^{1-x}\right) \right]\\ & = 
\frac{2 C_F}{2-x} \frac{\alpha_s(v^{1/(2-x)} Q)}{\alpha_s(Q)}\left[
(1-x)\left(\frac{5}{4}-\frac{\pi^2}{3}\right)-\int_0^1\!\! dz \frac{1+(1-z)^2}{z}\ln\left(z^{1-x}+(1-z)^{1-x}\right)
\right]\cF_\NLL(\lambda)\,.
\end{split}
\end{equation}
If we consider $\tau_x$ with respect to the WTA axis, we
obtain
\begin{equation}
  \label{eq:Frec-taux-WTA}
  \begin{split}
\dF{rec} & = \frac{2 C_F}{2-x} \frac{\alpha_s(v^{1/(2-x)} Q)}{\alpha_s(Q)}\left[
(1-x)\left(\frac{5}{4}-\frac{\pi^2}{3}\right)+\int_0^{1}\!\! dz \frac{1+(1-z)^2}{z}\ln\max[1-z,z]
\right]\cF_\NLL(\lambda) \\ & 
= \frac{2 C_F}{2-x} \frac{\alpha_s(v^{1/(2-x)} Q)}{\alpha_s(Q)}\left[
(1-x)\left(\frac{5}{4}-\frac{\pi^2}{3}\right)
+\left(\frac{3}{2}-\frac{\pi^2}{6} - \frac{3}{2}\ln 2\right)
\right]\cF_\NLL(\lambda)
\,.
  \end{split}
\end{equation}
Finally, for the moments of EEC we obtain
\begin{equation}
  \begin{split}
  \dF{rec} & = 
\frac{2 C_F}{2-x} \frac{\alpha_s(v^{1/(2-x)} Q)}{\alpha_s(Q)}\cF_\NLL(\lambda) \int_0^1\!\! dz \frac{1+(1-z)^2}{z} \ln\frac{[z(1-z)]^{1-x}}{z^{1-x}}\\ & 
=
2 C_F \frac{1-x}{2-x}\frac{\alpha_s(v^{1/(2-x)} Q)}{ \alpha_s(Q)}\left(\frac{5}{4}-\frac{\pi^2}{3}\right)\cF_\NLL(\lambda)\,.
\end{split}
\end{equation}
\subsection{Soft wide-angle corrections}
\label{sec:FC-wa-corr}

If we add to an ensemble of soft and collinear emissions,
$k_1,\dots,k_n$, a single soft emission $k$, at an angle $\theta$ with
respect to the thrust axis that is much larger than that of all other
emissions, we have
\begin{equation}
  \label{eq:k-wa}
  |\sin\theta_{k1}|\simeq |\sin\theta_{k2}|\simeq |\sin \theta|\,,
\end{equation}
where $\theta_{k\ell}$ is the angle between $k$ and $\tilde p_\ell$,
with $\ell=1,2$. Also, for any appropriate value of $x$, since $k$ is the emission
at the largest angle, its transverse momentum with respect to its 
emitter is the same as that with respect to the thrust axis. Therefore, for all considered observables, we have
\begin{equation}
  \label{eq:FC-wa}
  \begin{split}
    V_{\rm wa}(\ptilde,k,k_1,\dots,k_n) - V_{\rm sc}(\ptilde,k_1,\dots,k_n)& =\frac{k_t}{Q}\left(\frac{1+|\cos\theta|}{1-|\cos\theta|}\right)^{\frac{x-1}{2}} = \frac{k_t}{Q}e^{-(1-x)|\eta|} = V_{\rm sc}(k)\,.
  \end{split}
\end{equation}
Therefore,
$V_{\rm wa}(\ptilde,k,k_1,\dots,k_n)=V_{\mathrm{sc}}(\ptilde,k,k_1,\dots,k_n)$,
and $\dF{wa}=0$.

\subsection{Correlated corrections}
\label{sec:FC-correl-corr}

Since all the observables we consider are the same in the soft and
collinear limit, whenever we have any two soft emissions $k_a,k_b$,
collinear to the same leg $\ell$, together with an ensemble of soft-collinear
emissions $k_1,\dots,k_n$, we obtain
\begin{equation}
  \label{eq:FC-kakb}
  V_{\mathrm{sc}}(\ptilde,k_a,k_b,k_1,\dots,k_n) - V_{\mathrm{sc}}(\ptilde,k_1,\dots,k_n) = V_{\mathrm{sc}}(k_a)+V_{\mathrm{sc}}(k_b)\,.
\end{equation}
In terms of the variables defined in appendix~\ref{sec:2ps}, we have
\begin{equation}
  \label{eq:FC-rel}
  V_{\mathrm{sc}}(k_a)+V_{\mathrm{sc}} (k_b) = \left(\frac{\sqrt{k_t^2+m^2}}{Q} \right)^{x-1}e^{-(1-x)\eta^{(\ell)}}\left[z \left(\frac{|\vec q_a|}{Q}\right)^{2-x}+(1-z) \left(\frac{|\vec q_b|}{Q}\right)^{2-x}\right]\,.  
\end{equation}
Using the rescaled variables $\mu^2 \equiv m^2/k_t^2$ and
$\vec u_i \equiv \vec q_i/k_t$, and using the notation of section~\ref{sec:Fcorrel-additive}, we obtain
\begin{equation}
  \label{eq:fcorrel-FCx}
   f_{\rm correl}(z,\mu,\phi)\equiv (1+\mu^2)^{\frac{x-1}{2}}\,\tilde f_{\rm correl}(z,\mu,\phi;x)\,,
\end{equation}
with 
\begin{equation}
  \label{eq:ftilde-correl}
  \begin{split}
  &\tilde f_{\rm correl}(z,\mu,\phi;x) =z |\vec u_a|^{2-x}+(1-z) |\vec u_b|^{2-x} \\ & =
z \left(1+2\sqrt{\frac{1-z}{z}}\mu\cos\phi+\frac{1-z}{z}\mu^2\right)^{1-\frac{x}{2}}\!\!\! +
(1-z) \left(1-2\sqrt{\frac{z}{1-z}}\mu\cos\phi+\frac{z}{1-z}\mu^2\right)^{1-\frac{x}{2}}\,.  
  \end{split}
\end{equation}
Note that, for $x=0$, which is the same as $1-T$, we have
\begin{equation}
  \label{eq:fcorrel-beta2}
   f_{\rm correl}(z,\mu,\phi)   =
 \frac{1}{\sqrt{1+\mu^2}}
\left[z +(1-z)\mu^2+
(1-z) +z \mu^2\right] = \sqrt{1+\mu^2} \,.  
\end{equation}
This implies that, for $x=0$, one has
\begin{equation}
  V_{\rm sc}(k_a)+V_{\mathrm{sc}}(k_b) = 
  \frac{\sqrt{k_t^2+m^2}}{Q}e^{-\eta^{(\ell)}}=V(\ptilde,k_a+k_b)\,.
\end{equation}
Therefore, only for $x=0$ are the considered observables fully
inclusive with respect to multiple collinear splittings. This result
generalises to an arbitrary number of soft and collinear emissions.

\begin{figure}[htbp]
  \centering
  \includegraphics[width=.5\textwidth]{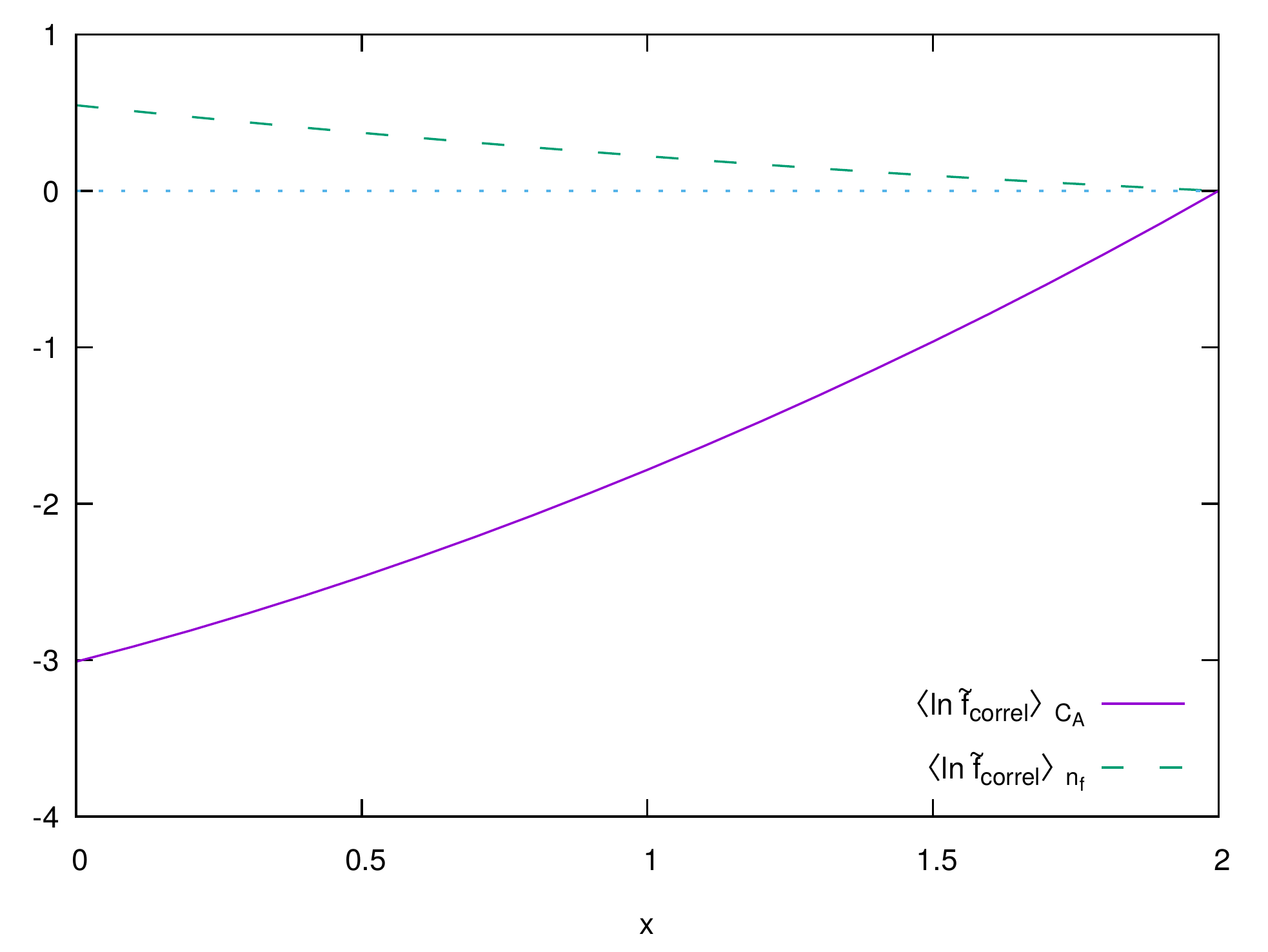}
  \caption{The corrections
    $\langle \ln\tilde f_{\rm correl}\rangle_{C_A}$ and
    $\langle \ln\tilde f_{\rm correl}\rangle_{n_f}$ as a function of
    $x$. }
  \label{fig:fcorrel}
\end{figure}

Now we can compute $\dF{correl}$ for any value of $x$ using the
general formula in Eq.~(\ref{eq:Fcorrel-additive-final}). We obtain
\begin{equation}
  \label{eq:FC-correl}
    \begin{split}
      \dF{correl}(\lambda) & = -\FNLL(\lambda)\frac{\lambda R^{\prime\prime}}{
        2\beta_0\alpha_s(Q)}
      \int_0^\infty\frac{d\mu^2}{\mu^2(1+\mu^2)} \int_0^1 dz
      \int_0^{2\pi}\frac{d\phi}{2\pi}\frac{1}{2!}C_{ab}(\mu,z,\phi)
\times      \\ & \qquad\qquad\qquad\qquad\qquad\qquad  \times 
\left[\frac{x-1}{2}\ln(1+\mu^2) +\ln\tilde f_{\rm
          correl}(\mu,z,\phi;x)\right] \\ & =-\FNLL(\lambda)\frac{\lambda R^{\prime\prime}}{
        2\beta_0\alpha_s(Q)} \left((1-x)\,(\pi\beta_0\zeta_2)+C_A \langle \ln\tilde f_{\rm correl}\rangle_{C_A} + n_f \langle \ln\tilde f_{\rm correl}\rangle_{n_f} \right) \,,
   \end{split}
 \end{equation}
where 
\begin{equation}
  \label{eq:fcorrel-average}
  \begin{split}
  \langle \ln\tilde f_{\rm correl}\rangle_{C_A} & = \int_0^\infty\frac{d\mu^2}{\mu^2(1+\mu^2)} \int_0^1 dz
      \int_0^{2\pi}\frac{d\phi}{2\pi}\frac{1}{2!} \left(2 \mathcal{S}+\mathcal{H}_g\right)\ln\tilde f_{\rm
          correl}(\mu,z,\phi;x)\,, \\
  \langle \ln\tilde f_{\rm correl}\rangle_{n_f} & = \int_0^\infty\frac{d\mu^2}{\mu^2(1+\mu^2)} \int_0^1 dz
      \int_0^{2\pi}\frac{d\phi}{2\pi}\frac{1}{2!} \mathcal{H}_q\ln\tilde f_{\rm
          correl}(\mu,z,\phi;x)\,.
  \end{split}
\end{equation}
In the above equation, $2\mathcal S$ and $\mathcal{H}_g$ are defined
in Eqs.~(\ref{eq:2S}) and~(\ref{eq:Hg}) respectively, and
$\mathcal{H}_g$ is defined in Eq.~(\ref{eq:Hq}).  We have computed
$\langle \ln\tilde f_{\rm correl}\rangle_{C_A}$ and
$\langle \ln\tilde f_{\rm correl}\rangle_{n_f}$ numerically as a
function of $x$, and the result can be found in
Fig.~\ref{fig:fcorrel}.
For $x=0$, $\langle \ln\tilde f_{\rm correl}\rangle_{C_A}$ and
$\langle \ln\tilde f_{\rm correl}\rangle_{n_f}$ can be computed
analytically, which gives
 \begin{equation}
   \label{eq:vev-lnfcorrel}
   \langle\ln\tilde f_{\rm correl}\rangle_{C_A} = -\frac{11  \zeta_2}{6}\,,\qquad
   \langle\ln\tilde f_{\rm correl}\rangle_{n_f} = \frac{\zeta_2}{3} \,.
 \end{equation}

\subsection{Matching and issues with Sudakov shoulders for $FC_x$ }
It is interesting to study the matching to fixed order for the moments
of energy energy correlation~\eqref{eq:FC-def}. Such observables
feature a Sudakov shoulder~\cite{Catani:1997xc}, whose position can
get dangerously close to the Sudakov peak for certain values of
$x$. To examine this feature we now match the resummed NNLL
distributions to NLO fixed-order differential cross sections obtained
with {\tt EVENT2}~\cite{Catani:1996vz}. Although we only analyse
$FC_x$ below, the procedure discussed in the following applies to all
observables considered in this article. The matching is performed
according to the log-R scheme (see for
instance~\cite{Catani:1992ua,Monni:2011gb}).
As it is customary in resummed calculations, to probe the size of
subleading logarithmic terms we introduce a rescaling constant $x_V$
as \begin{equation}
\ln\frac{1}{v} = \ln\frac{x_V}{v}-\ln x_V\,,
\end{equation}
and expand the cross section around $\ln x_V/v$ neglecting subleading
terms.\footnote{For details about how the resummed formula and the
  expansion coefficients change see e.g.~Ref.~\cite{Monni:2011gb}
  where one has to replace $\ln x_L\rightarrow -\ln x_V$.}
Eventually we modify the resummed logarithm $\ln x_V/v$ in order to
impose that the total cross section is reproduced at the kinematical
endpoint $v_{\rm max}$ 
\begin{equation}
\label{eq:mod-logs}
\ln\frac{x_V}{v} \rightarrow \frac{1}{p}\ln\left(1+\left(
    \frac{x_V}{v}\right)^p - \left( \frac{x_V}{v_{\rm max}}\right)^p
\right)\,. 
\end{equation}
Here, $p$ denotes a positive number which controls how quickly the
logarithms are switched off close to the endpoint. Since in the
following we do not perform a phenomenological study, we simply set
$p=1$ and $v_{\rm max}=1$ for the sake of simplicity.

To obtain our central predictions we set $\mu_R = Q$, with $Q$ being
the centre-of-mass energy of the hard scattering, corresponding to
$\alpha_s(M_Z) = 0.118$, and $x_V=1$.  We then construct the
uncertainty bands by varying $\mu_R$ and $x_V$ individually by a
factor of two in either direction. The relevant formulae for the scale
dependence are reported in the appendix of Ref.~\cite{Banfi:2014sua}.

\begin{figure}[!ht]
  \centering
  \includegraphics[width=0.45\columnwidth]{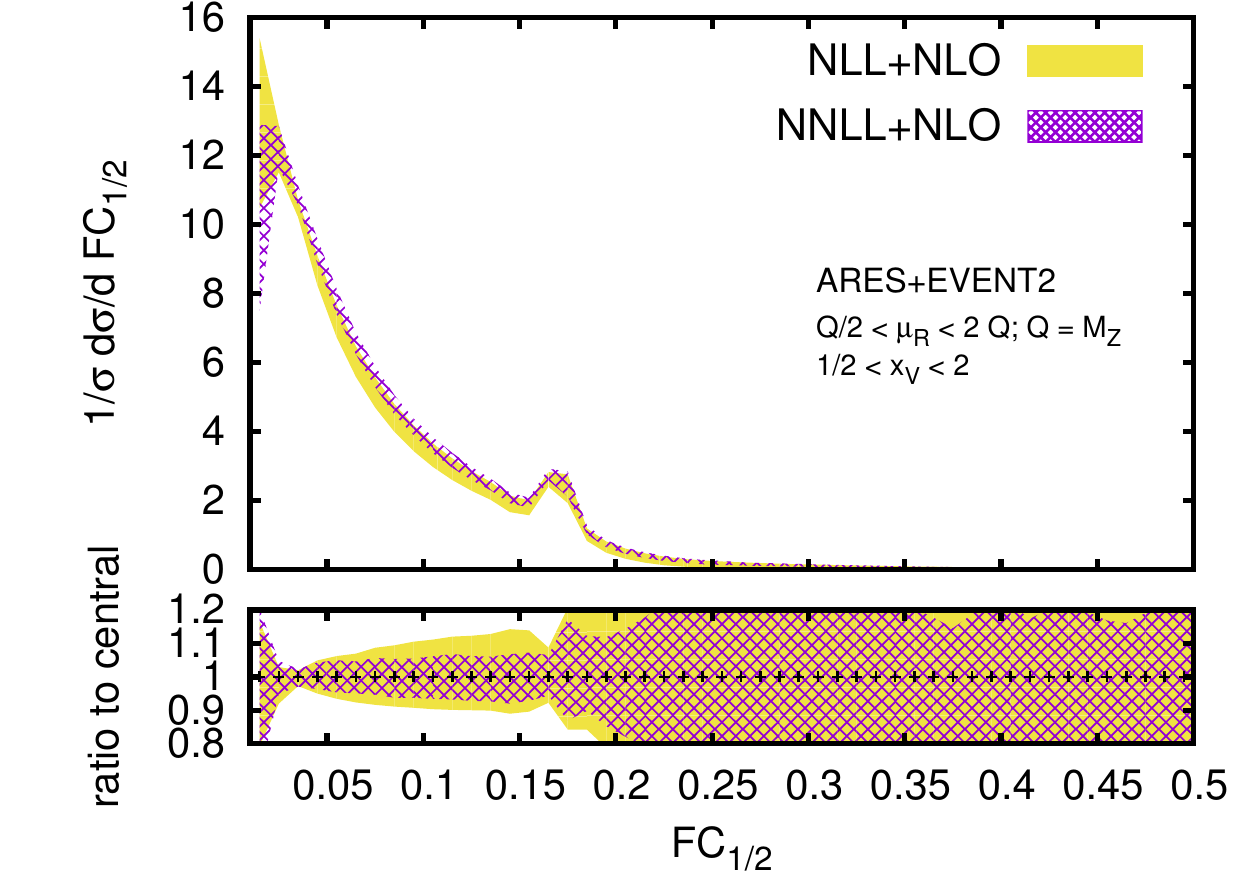} 
  \includegraphics[width=0.45\columnwidth]{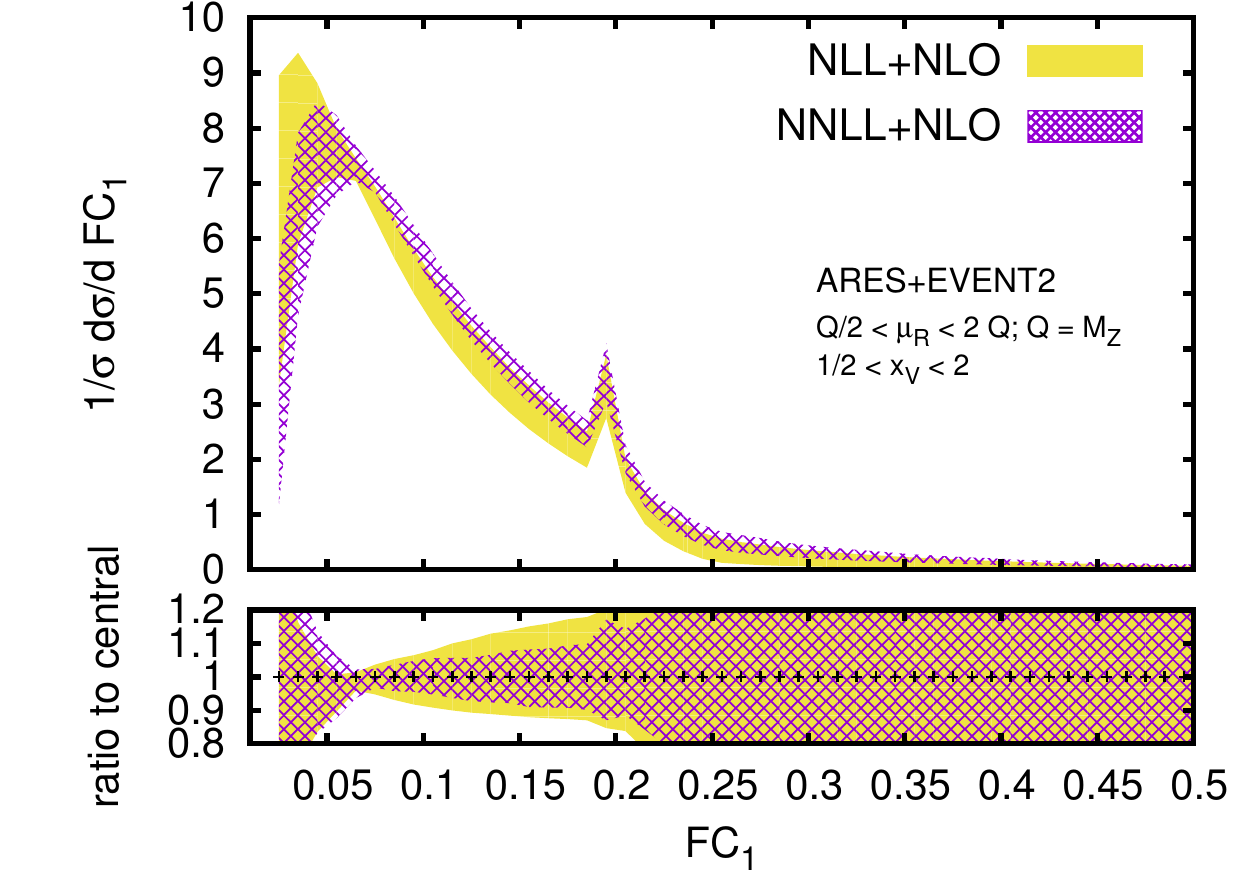} 
  \caption{NNLL+NLO and NLL+NLO distributions for the two moments of
    EEC $FC_{1/2}$ and $FC_1$ at $Q=M_Z$. The lower inset shows the ratio to the
    relative central value for each of the two bands.}
  \label{fig:MZ}
\end{figure}
Figure~\ref{fig:MZ} shows the NLO differential distribution matched to
both NLL and NNLL for the cases $x=1/2$ and $x=1$. From the plots one
can appreciate the following two interesting features.

The first is that the size of NNLL corrections increases with
$x$. This can be explained by inspecting the parametrisation of the
observable in the soft and collinear limit~\eqref{eq:Vsc-k}. The
characteristic transverse momentum of the soft radiation is
$k_t\sim Q v$, while the hard-collinear radiation occurs at scales
$k_t\sim Q v^{\frac{1}{2-x}}$, with $x<2$.
The soft scale is therefore lower than the collinear scales for
$x < 1$, the two coincide for $x=1$, and the situation is inverted for
$2 > x > 1$. For a given value $v$ of the observable, the typical size
of the soft logarithms (and hence of the soft corrections) does not
depend on the moment parameter $x$. Conversely, the size of the
hard-collinear logarithms increases with $x$, hence leading to larger
subleading corrections. One also expects that corrections beyond NNLL
become more sizeable as $x$ increases, as it is reflected by the scale
uncertainty band in Figure~\ref{fig:MZ}. For $x>1$ the subleading
corrections grow very large as the collinear scale becomes smaller
than the soft one, which corresponds to a badly convergent logarithmic
series. A consequence of this fact is that for $x>1$ the abscissa of
the Landau pole moves towards larger values of $v$, and hence the
differential distribution becomes non-perturbative at moderate values
of $v$.

\begin{figure}[h!]
  \centering
  \includegraphics[width=0.45\columnwidth]{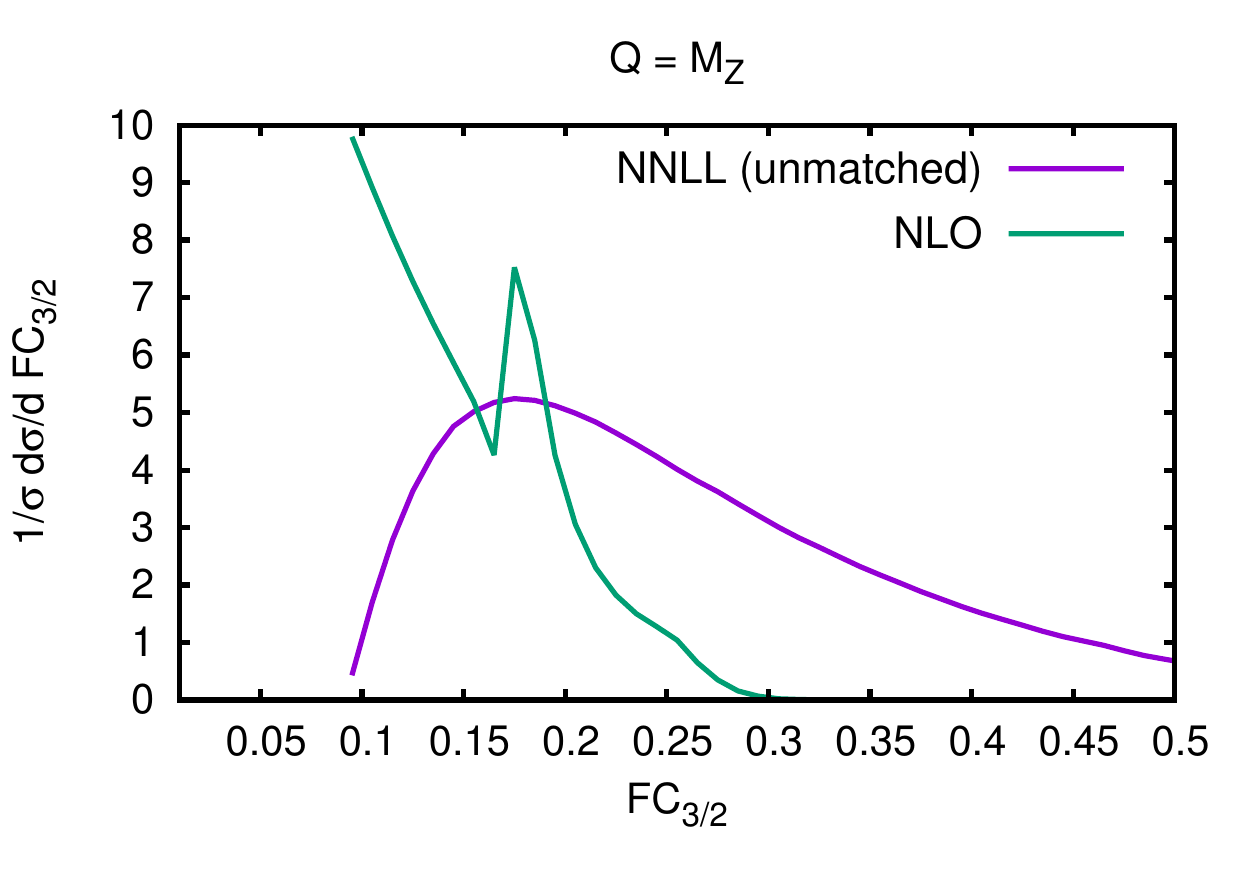} 
  \includegraphics[width=0.45\columnwidth]{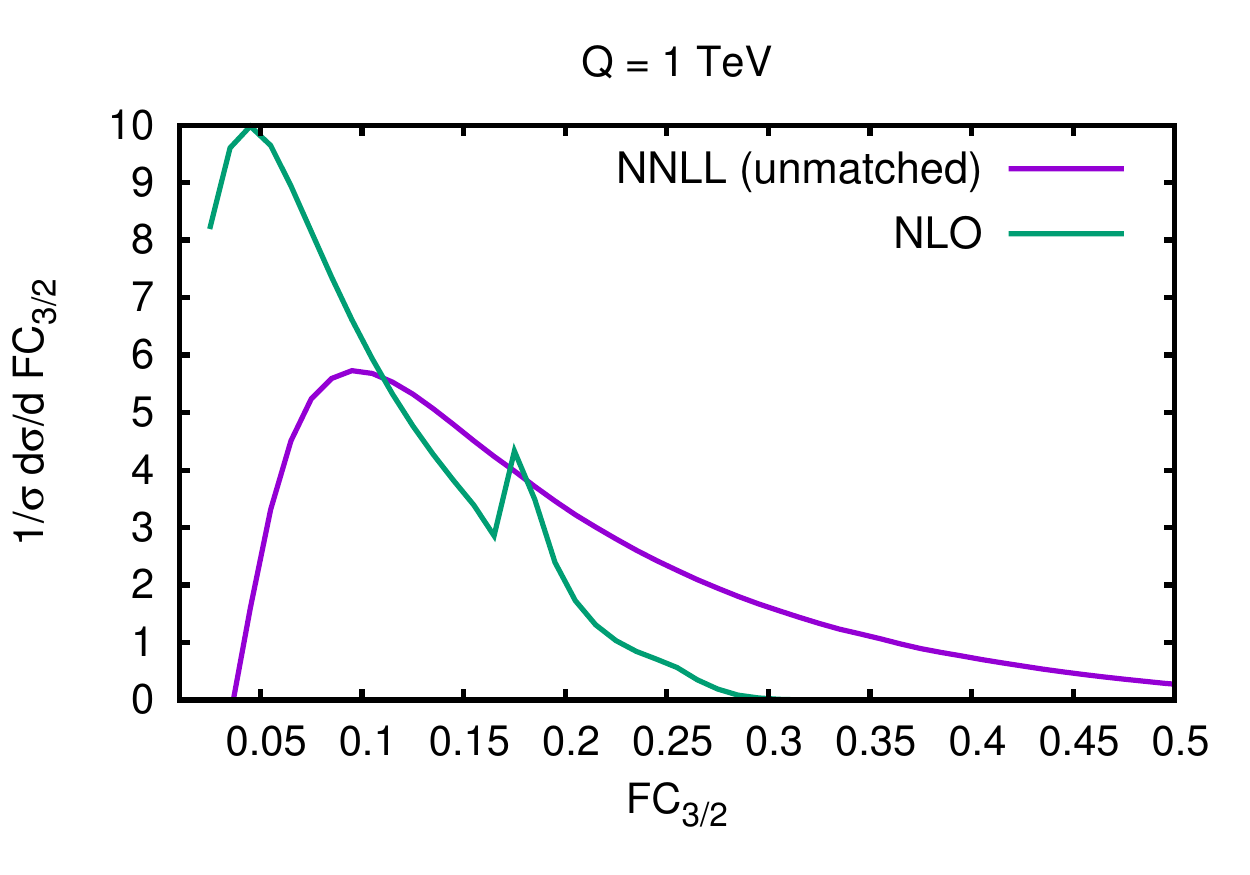} 
  \caption{Comparison between the unmatched NNLL distribution with
    central scales and the NLO prediction for the moment of EEC
    $FC_{3/2}$ at $Q=M_Z$ (left) and $Q=1$~TeV (right).}
  \label{fig:shoulders}
\end{figure}

A second interesting observation is that the relative distance between
the Sudakov peak and the shoulder decreases for increasing $x$. This
implies that there is a value of $x$ for which the two overlap. Such a
situation can be observed in the left plot of
Figure~\ref{fig:shoulders} for $x=3/2$, where the curves are obtained
at the $Z$ resonance and for central values of the scales.

In this case the solution provided by the resummation in the two-jet
limit is obviously unphysical. The position of the Sudakov peak
represents the bulk of the soft and collinear radiation probability in
the two-jet configuration, and it coincides with the kinematic
endpoint for the three-jet configuration, above which the distribution
is again dominated by soft and collinear emissions.
 
This phenomenon is due to the violation of momentum conservation in
the formulation of the resummed calculation, in which the exact
kinematics in the presence of an extra hard parton is ignored. In such
a situation one should perform a simultaneous resummation of the
Sudakov logarithms treated here together with the logarithms that
originate at the shoulder. This is currently out of reach at the
logarithmic order analysed in this article.

\begin{figure}[h!]
  \centering
  \includegraphics[width=0.45\columnwidth]{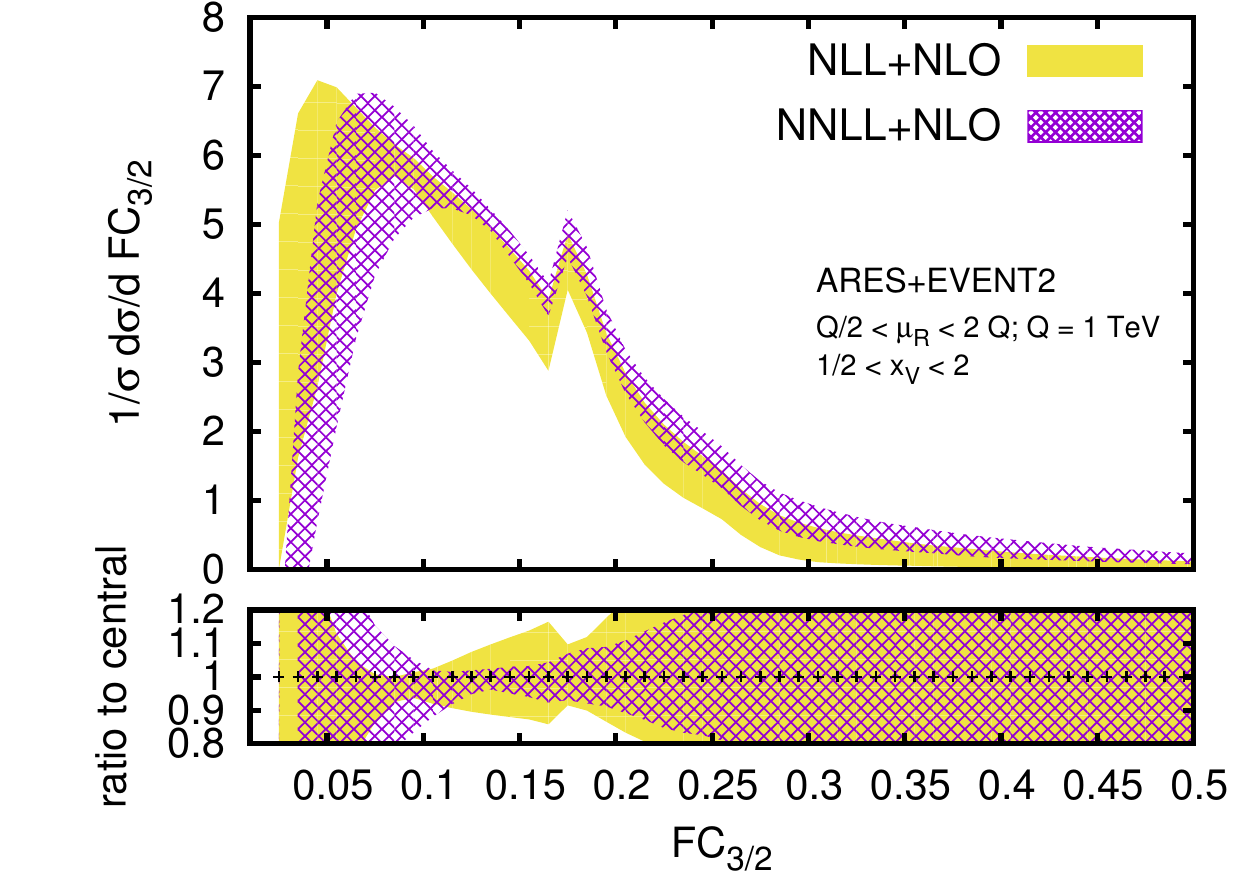} 
  \caption{NNLL+NLO and NLL+NLO distributions for the moment of EEC
    $FC_{3/2}$ at $Q=1$~TeV. The lower inset shows the ratio to the
    relative central value.}
  \label{fig:1TeV}
\end{figure}

While in this case a matching to fixed order results in an unphysical
prediction, the right plot of Figure~\ref{fig:shoulders} shows that the
situation improves at higher collider energies.
As can be seen from this plot, for higher collider energies the
position of the Sudakov peak moves towards smaller values of the
observable, driven by the smaller coupling constant, while the
position of the shoulder does note depend on $Q$. At these scales the
resummed result is physical and can be matched to the fixed order. An
example is reported in Figure~\ref{fig:1TeV}, where the matched
distribution for $Q=1$~TeV is shown.

\section{Conclusions}
\label{sec:the-end}

In this paper we have completed the study of jet observables at NNLL
accuracy in $e^+ e^-$ annihilation, that started in
refs.~\cite{Banfi:2014sua,Banfi:2016zlc}. These results constitute the
core of the \texttt{ARES} method for the semi-numerical resummation of
jet observables at NNLL accuracy, which generalises the NLL procedure
of refs.~\cite{Banfi:2004yd,Banfi:2003je,Banfi:2001bz}. This involves
the calculation of an observable dependent Sudakov form factor, the
radiator, which encodes the all-order cancellation of infrared
singularities between real and virtual contributions, and that we have
computed at NNLL accuracy for a generic rIRC safe observable.

As a byproduct, we have defined a generalisation of the well known CMW
physical coupling in the soft limit, and given a closed expression for
its relation to the $\overline{\rm MS}$ coupling up to
${\cal O}(\alpha_s^3)$. This quantity is a universal ingredient for
all resummations of rIRC safe observables, and as such it constitutes
one of the main ingredients for a NNLL accurate parton shower
algorithm.

As an application, we have computed NNLL resummed distributions for
angularities and fractional moments of EEC, for all allowed values of
the parameter $x$ they depend on. We have also presented a very basic
phenomenology of moments of EEC, highlighting their main features. A
particularly severe issue is that, for $x>1$, the Sudakov peak of the
differential distribution, where the observable should be dominated by
multiple soft-collinear emissions, becomes dangerously close to the
edge of the phase space for real emissions, so that the core
approximation underlying soft-collinear resummations breaks down. This
situation is severe at LEP energies and prevents us to make any sense
of resummed prediction matched to fixed order for certain values of
$x>1$. This is not the case at a future $e^+e^-$ collider with
centre-of-mass energy of $1\,\mathrm{TeV}$, where the Sudakov peak
moves towards lower values of the observable, while the position of
the kinematical boundary stays unchanged. This is a feature that
should be considered when using these observables for phenomenology.

We stress that the procedure outlined in this paper is only an example
on how to construct a Sudakov radiator. The same calculation could for
instance be performed at all orders using the methods of
Soft-Collinear Effective Theory (SCET), along the lines of what has
been shown for thrust in Ref.~\cite{Bauer:2018svx}. To achieve a full
resummation, the radiator should be supplemented by appropriate NLL
and NNLL corrections due to the resolved real radiation. In this
paper, we have defined the Sudakov radiator in such a way that all
NLL, and most NNLL corrections are the same as in
refs.~\cite{Banfi:2014sua,Banfi:2016zlc}. The only exception is the
NNLL correction $\delta \mathcal{F}_{\rm correl}$, which we had to
redefine to ensure that the radiator could be computed analytically
for a generic rIRC safe observable.

We also note that this very same radiator could in principle be used
also for processes with incoming hadrons, for reactions with two hard
emitting legs at the Born level. In that case, all corrections due to
soft radiation, $\mathcal{F}_{\NLL},\dF{sc},\dF{wa},\dF{correl}$ stay
unchanged, and one has to evaluate parton distribution functions at
the factorisation scales of order $v^{1/(a+b_\ell)}Q$, and recompute
only the hard-collinear contributions
$C^{(1)}_{{\rm hc},\ell},\dF{hc},\dF{rec}$, as done for instance in
Ref.~\cite{Monni:2016ktx,Bizon:2017rah} for certain classes of
observables. For processes with more than two legs, all terms in the
master formula~\eqref{eq:Sigma-NNLL} must be redefined in order to
account for the structure of the wide-angle soft radiation. This will
be left for future work.

In summary, this work completes the formulation of a general method
for to the calculation of any jet observable in processes with two
legs, that can be systematically generalised to more complicated
cases. We hope that the results presented here will define a solid
starting point for future systematic studies of jet observables at all
perturbative orders.

\section*{Acknowledgements}

We are grateful to G.~Salam for helpful discussions on the topics
discussed in this article, and to M.~Procura for a cross check of the
results for WTA angularities. The work of A.B. and B.K.E. is supported
by the Science Technology and Facilities Council (STFC) under grant
number ST/P000819/1. The work of P.F.M.\ has been supported by a Marie
Sk\l{}odowska Curie Individual Fellowship of the European Commission's
Horizon 2020 Programme under contract number 702610 Resummation4PS.

\appendix
\section{Correlated two-parton emission}
\label{sec:2ps}

We start by decomposing the momenta of the two partons $k_a$ and $k_b$
as in Eq.~\eqref{eq:sudakov}.  We then introduce relative
variables to parameterise the two-parton phase space, as follows
\begin{equation}
  \label{eq:2vars}
  \begin{split}
    &  z^{(\ell)}_a=z\, z^{(\ell)}\,,\quad z_b^{(\ell)}=(1-z)\,z^{(\ell)}\,,\\
    &\vec q_a = \frac{\vec k_{ta}}{z}\,,\quad \vec q_b = \frac{\vec
      k_{tb}}{1-z}\,,
  \end{split}
\end{equation}
in terms of which the Lorentz invariant phase-space in $4-2\epsilon$
dimensions becomes
\begin{equation}
  \label{eq:ps2-take1}
  [dk_a][dk_b]=\frac{1}{(4\pi)^2}\frac{dz^{(\ell)}}{z^{(\ell)}} dz [z(1-z)]^{1-2\epsilon} \frac{d^{2-2\epsilon}q_a}{(2\pi)^{2-2\epsilon}} \frac{d^{2-2\epsilon}q_b}{(2\pi)^{2-2\epsilon}} \,. 
\end{equation}
Another useful change of variables is
\begin{equation}
  \label{eq:2vars-take2}
  \vec k_t = \vec k_{ta}+\vec k_{tb}\,,\qquad \vec q = \vec q_a-\vec q_b\,,
\end{equation}
in terms of which the phase-space becomes
\begin{equation}
  \label{eq:ps2-take2}
  [dk_a][dk_b]=\frac{1}{(4\pi)^2}\frac{dz^{(\ell)}}{z^{(\ell)}} \frac{d^{2-2\epsilon}k_t}{(2\pi)^{2-2\epsilon}} dz [z(1-z)]^{1-2\epsilon}  \frac{d^{2-2\epsilon}q}{(2\pi)^{2-2\epsilon}} = [dk] \frac{dz [z(1-z)]^{1-2\epsilon}}{4\pi}  \frac{d^{2-2\epsilon}q}{(2\pi)^{2-2\epsilon}}\,,
\end{equation}
where we have been able to factor out the phase space $[dk]$ defined
in Eq.~(\ref{eq:dk-final}). Last, one can isolate the integration over
$\phi$, the angle between $\vec k_t$ and $\vec q$, and introduce
\begin{equation}
  \label{eq:m2}
  m^2\equiv (k_a+k_b)^2=z(1-z)q^2\,,
\end{equation}
to obtain yet another expression for the two-body phase space
\begin{equation}
  \label{eq:ps2-take3}
    [dk_a][dk_b]=[dk] \frac{dz
      [z(1-z)]^{-\epsilon}}{(4\pi)^2}\frac{dm^2}{(m^2)^\epsilon} \frac{d\Omega_{2-2\epsilon}}{(2\pi)^{1-2\epsilon}}\,.
\end{equation}
The factor $d\Omega_{2-2\epsilon}$ is the azimuthal phase space for
the vector $\vec q$ with respect to $\vec k_t$. Explicitly, this is given by
\begin{equation}
  \label{eq:azimuthal-ps}
  d\Omega_{2-2\epsilon} = \frac{(4\pi)^{\epsilon}}{\sqrt{\pi}\Gamma(\frac{1}{2}-\epsilon)} d\phi (\sin^2\phi)^{-\epsilon}\,.
\end{equation}
 where the relative
angle $ \phi$ in the range $0<\phi<\pi$.

In terms of these variables, the correlated matrix element
$\tilde M_{s,0}^2(k_a,k_b)$ is given by
\begin{equation}
  \label{eq:tildeM2}
  \tilde M_{s,0}^2(k_a,k_b) =(4\pi\alpha_s \mu_R^{2\epsilon})^2\frac{8 C_\ell}{m^2(m^2+k_t^2)} C_{ab}(k_a,k_b)
\,,
\end{equation}
where $\mu_R$ is the renormalisation scale, $C_\ell$ is the colour
factor associated with the emitting leg, and
\begin{equation}
\label{eq:Corr}
  C_{ab}(k_a,k_b) = C_A(2 \mathcal{S}+\mathcal{H}_g) + n_f \mathcal{H}_q \,.
 \end{equation}
 The contribution due to two final-state quarks in Eq.~\eqref{eq:Corr}
 has been multiplied by two, to compensate for the overall $1/2!$
 factor in Eq.~\eqref{eq:Fcorrel-new}.  The three functions
 $\mathcal{S}$, $\mathcal{H}_g$ and $\mathcal{H}_q$ are the
 $4-2\epsilon$-dimensional counterparts of the homonymous terms
 defined in Ref.~\cite{Dokshitzer:1997iz}. They depend only on the
 dimensionsless variables $z$, $ \phi$ and $\mu^2\equiv m^2/k_t^2$. It
 is also useful to introduce the rescaled momenta
 $\vec u_i= \vec q_i/k_t$, such that
 \begin{equation}
   \label{eq:xqi}
   u_a^2 = 1+2\sqrt{\frac{1-z}{z}}\mu\cos \phi + \frac{1-z}{z}\mu^2\,,\qquad 
   u_b^2 = 1-2\sqrt{\frac{z}{1-z}}\mu\cos \phi + \frac{z}{1-z}\mu^2\,.
 \end{equation}
In terms of these variables, we have
 \begin{subequations}
\begin{align}
  \label{eq:2S}
  2\mathcal{S} & = \frac{1}{z(1-z)}\left[\frac{1-(1-z)\mu^2/z}{u_a^2}+\frac{1-z\mu^2/(1-z)}{u_b^2}\right] \\
  \label{eq:Hg}
  \mathcal{H}_g & =-4+(1-\epsilon)\frac{z(1-z)}{1+\mu^2}\left(2\cos  \phi+\frac{(1-2z)\mu}{\sqrt{z(1-z)}}\right)^2\nonumber \\ & +\frac{1}{2(1-z)}\left[1-\frac{1-(1-z)\mu^2/z}{u_a^2}\right] 
+\frac{1}{2z}\left[1-\frac{1-z\mu^2/(1-z)}{u_b^2}\right]
\\
  \label{eq:Hq}
  \mathcal{H}_q& =1-\frac{z(1-z)}{1+\mu^2}\left(2\cos  \phi+\frac{(1-2z)\mu}{\sqrt{z(1-z)}}\right)^2\,.
\end{align}
 \end{subequations}
 Note that, in the limit $\mu^2\to 0$, one recovers the azimuthally
 unaveraged splitting functions, in particular
 \begin{subequations}
   \begin{align}
     & 2\mathcal{S}+\mathcal{H}_g \to 2\left[\frac{1}{z(1-z)}-2+2(1-\epsilon)z(1-z)\cos^2 \phi\right]\,,\\
     & \mathcal{H}_q \to  1-4 z(1-z)\cos^2 \phi\,. 
   \end{align}
 \end{subequations}

\end{document}